\documentclass[aps, prx, reprint, showpacs, superscriptaddress, longbibliography, floatfix]{revtex4-2}

\usepackage[english]{babel}
\usepackage[bbgreekl]{mathbbol}
\usepackage{amsfonts}

\DeclareSymbolFontAlphabet{\mathbb}{AMSb}
\DeclareSymbolFontAlphabet{\mathbbl}{bbold}

\usepackage{enumitem}
\usepackage{autofigs9}
\usepackage{braket}
\usepackage{amsmath,amssymb}
\usepackage{linenoAMS}
\usepackage{txfonts}
\usepackage{comment}
\usepackage{bm}
\usepackage{mathtools}

\graphicspath{{figures/}}
\usepackage[usenames, dvipsnames, svgnames, table]{xcolor}
\usepackage[colorlinks=true, citecolor=RoyalBlue,
            linkcolor=BrickRed, urlcolor=ForestGreen]{hyperref}
\usepackage[capitalise]{cleveref}

\usepackage{graphicx}
\usepackage{xfrac}
\usepackage{ulem}
\usepackage{cancel}

\newcommand{\mat}[1]{\ensuremath{\mathbf{#1}}}
\newcommand{\Tmat}[1]{\ensuremath{\mathbbl{#1}}}
\newcommand{\Tvec}[1]{\overset{\raisebox{-2pt}{\text{\tiny$\bm\rightarrow$}}}{\mathbbl{#1}}}
\newcommand{\Marrowbbl}[1]{\overset{\raisebox{-1pt}{\text{\tiny$\bm\Leftrightarrow$}}}{#1}}
\newcommand{\Dmat}[1]{\ensuremath{\Marrowbbl{\mathbf{#1}}}}
\newcommand{\Mrarrowbbl}[1]{\overset{\raisebox{-1pt}{\text{\tiny$\bm\Rightarrow$}}}{#1}}
\newcommand{\Dvec}[1]{\ensuremath{\Mrarrowbbl{\mathbf{#1}}}}
\newcommand{\delay}{\ensuremath{L}}
\newcommand{\K}{\ensuremath{\mathcal{K}}}

\newcommand{\conj}[1]{\ensuremath{{#1}^*}}

\newcommand{\MM}{\ensuremath{\Upsilon}}
\newcommand{\MMR}{\ensuremath{\Upsilon_{\text{R}}}}
\newcommand{\MMI}{\ensuremath{\Upsilon_{\text{I}}}}
\newcommand{\MMO}{\ensuremath{\Upsilon_{\text{O}}}}
\newcommand{\MMF}{\ensuremath{\Upsilon_{\text{F}}}}
\newcommand{\cMM}{\ensuremath{1{-}\MM}}
\newcommand{\cMMR}{\ensuremath{1{-}\MMR}}
\newcommand{\cMMI}{\ensuremath{1{-}\MMI}}
\newcommand{\cMMO}{\ensuremath{1{-}\MMO}}
\newcommand{\cMMF}{\ensuremath{1{-}\MMF}}

\newcommand{\sqzang}{\ensuremath{\phi}}

\newcommand{\aFactor}{\ensuremath{\alpha}}
\newcommand{\bFactor}{\ensuremath{\beta}}
\newcommand{\psiR}{\ensuremath{\psi_\text{R}}}
\newcommand{\psiG}{\ensuremath{\psi_\text{G}}}
\newcommand{\psiS}{\ensuremath{\psi_\text{S}}}
\newcommand{\psiI}{\ensuremath{\psi_\text{I}}}
\newcommand{\psiO}{\ensuremath{\psi_\text{O}}}
\newcommand{\psiF}{\ensuremath{\psi_\text{F}}}

\newcommand{\Ophi}{\ensuremath{\phi}}
\newcommand{\phiRMS}{\ensuremath{\phi_{\text{rms}}}}
\newcommand{\phiRMSsq}{\ensuremath{\phi^2_{\text{rms}}}}

\newcommand{\thetaRMSsq}{\ensuremath{\theta^2_{\text{rms}}}}

\newcommand{\TLO}{\ensuremath{\Tvec{v}^{\dagger}}}
\newcommand{\TLOa}{\ensuremath{\Tvec{v}}}
\newcommand{\DLO}{\ensuremath{\Dvec{v}^\dagger}}

\newcommand{\Hloss}{\ensuremath{\Tmat{H}_{\text{loss}}}}

\newcommand{\Mq}{\ensuremath{m_q}}
\newcommand{\Mp}{\ensuremath{m_p}}
\newcommand{\cMp}{\ensuremath{\conj{m}_p}}

\newcommand{\sql}{\ensuremath{\text{sql}}}
\newcommand{\subI}{\ensuremath{\text{I}}}
\newcommand{\subO}{\ensuremath{\text{O}}}
\newcommand{\subR}{\ensuremath{\text{R}}}
\newcommand{\subFC}{\ensuremath{\text{FC}}}
\newcommand{\subfc}{\ensuremath{\text{fc}}}
\newcommand{\IRO}{\ensuremath{\text{IRO}}}

\newcommand{\subF}{\ensuremath{\text{F}}}

\newcommand{\arm}{\ensuremath{\text{a}}}
\newcommand{\src}{\ensuremath{\text{s}}}
\newcommand{\Arm}{\ensuremath{\text{A}}}
\newcommand{\Src}{\ensuremath{\text{S}}}
\newcommand{\itm}{\ensuremath{\text{a}}}
\newcommand{\etm}{\ensuremath{\text{e}}}
\newcommand{\srm}{\ensuremath{\text{s}}}

\newcommand{\tfh}{\ensuremath{\mathfrak{h}}}
\newcommand{\tfr}{\ensuremath{\mathfrak{r}}}
\newcommand{\tfrH}{\ensuremath{\mathfrak{r}_{\text{hom}}}}

\newcommand{\smu}{\ensuremath{{\,\mu}}}
\newcommand{\snu}{\ensuremath{{\,\nu}}}
\begin{document}

\title{LIGO's Quantum Response to Squeezed States}

\overfullrule 0pt 
\parskip0pt
\hyphenpenalty9999

\author{L.~McCuller}
\email{lee.mcculler@ligo.org}  
\affiliation{LIGO, Massachusetts Institute of Technology, Cambridge, MA 02139, USA}

\author{S.~E.~Dwyer}  
\affiliation{LIGO Hanford Observatory, Richland, WA 99352, USA}

\author{A.~C.~Green}
\affiliation{University of Florida, Gainesville, FL 32611, USA}

\author{Haocun~Yu}  
\affiliation{LIGO, Massachusetts Institute of Technology, Cambridge, MA 02139, USA}

\author{L.~Barsotti}  
\affiliation{LIGO, Massachusetts Institute of Technology, Cambridge, MA 02139, USA}

\author{C.~D.~Blair}  
\affiliation{LIGO Livingston Observatory, Livingston, LA 70754, USA}

\author{D.~D.~Brown}
\affiliation{OzGrav, University of Adelaide, Adelaide, South Australia 5005, Australia}

\author{A.~Effler}  
\affiliation{LIGO Livingston Observatory, Livingston, LA 70754, USA}

\author{M.~Evans}  
\affiliation{LIGO, Massachusetts Institute of Technology, Cambridge, MA 02139, USA}

\author{A.~Fernandez-Galiana}  
\affiliation{LIGO, Massachusetts Institute of Technology, Cambridge, MA 02139, USA}

% ...


\begin{abstract}
Gravitational Wave interferometers achieve their profound sensitivity by combining a Michelson interferometer with optical cavities, suspended masses, and now, squeezed quantum states of light. These states modify the measurement process of the LIGO, VIRGO and GEO600 interferometers to reduce the quantum noise that masks astrophysical signals; thus, improvements to squeezing are essential to further expand our gravitational view of the universe. Further reducing quantum noise will require both lowering decoherence from losses as well more sophisticated manipulations to counter the quantum back-action from radiation pressure. Both tasks require fully understanding the physical interactions between squeezed light and the many components of km-scale interferometers. To this end, data from both LIGO observatories in observing run three are expressed using frequency-dependent metrics to analyze each detector's quantum response to squeezed states. The response metrics are derived and used to concisely describe physical mechanisms behind squeezing's simultaneous interaction with transverse-mode selective optical cavities and the quantum radiation pressure noise of suspended mirrors. These metrics and related analysis are broadly applicable for cavity-enhanced optomechanics experiments that incorporate external squeezing, and -- for the first time -- give physical descriptions of every feature so far observed in the quantum noise of the LIGO detectors.
\end{abstract}
\maketitle

\section{Introduction}
The third observing run of the global gravitational wave network has not only produced a plethora of varied and unique astrophysics events \cite{Abbott-AAP20-PopulationProperties, Abbott-AAP20-GWTC2Compact}, it has defined a milestone in quantum metrology: that the LIGO, VIRGO and GEO600 observatories are now all reliably improving their scientific output by incorporating squeezed quantum states \cite{Tse-PRL19-QuantumEnhancedAdvanced, VirgoCollaboration-PRL19-IncreasingAstrophysical, Lough-PRL21-FirstDemonstration}. This marks the transition where optical squeezing, a widely researched, emerging quantum technology, has become an essential component producing new observational capability.

For advanced LIGO, observing run three provides the first peek into the future of quantum enhanced interferometry, revealing challenges and puzzles to be solved in the pursuit of ever more squeezing for ever greater observational range. Studying quantum noise in the LIGO interferometers is not simple. The audio-band data from the detectors contains background noise from many optical, mechanical and thermal sources, which must be isolated from the purely quantum contribution that responds to squeezing. All the while, the interferometers incorporate optical cavities, auxiliary optical fields, kg-scale suspended optics, and radiation pressure forces. The background noise and operational stability of the LIGO detectors is profoundly improved in observing run three \cite{Buikema-PRD20-SensitivityPerformance}, enabling new precision observations of the interactions between squeezed states and the complex optomechanical detectors.

Quantum radiation pressure noise (QRPN) is the most prominent new observation from squeezing \cite{Yu-N20-QuantumCorrelations, Acernese-PRL20-QuantumBackaction}. QRPN results from the coupling of photon momentum from the amplitude quadrature of the light into the phase quadrature, as radiation force fluctuation integrates into mirror displacement uncertainty. When vacuum states enter the interferometer, rather than squeezed states, QRPN imposes the so-called standard quantum limit
\cite{Braginsky-S80-QuantumNondemolition, Braginsky-92-QuantumMeasurement, Braginsky-RMP96-QuantumNondemolition},
bounding the performance of GW interferometers. Because the QRPN coupling between quadratures is coherent, squeezed states allow the SQL to be surpassed
\cite{Kimble-PRD01-ConversionConventional, Yu-N20-QuantumCorrelations}.
Both surpassing the SQL and increasing the observing range is possible by using a frequency-dependent squeezing (FDS) source implemented with a quantum filter cavity
\cite{Kimble-PRD01-ConversionConventional, McCuller-PRL20-FrequencyDependentSqueezing, Zhao-PRL20-FrequencyDependentSqueezed, Oelker-PRL16-AudioBandFrequencyDependent, Chelkowski-PRA05-ExperimentalCharacterization, Whittle-PRD20-OptimalDetuning, Khalili-PRD10-OptimalConfigurations, Evans-PRD13-RealisticFilter, Kwee-PRD14-DecoherenceDegradation}. LIGO is including such a source in the next observing run as part of its ``A+'' upgrade\cite{Whittle-PRD20-OptimalDetuning, McCuller-PRL20-FrequencyDependentSqueezing}. To best utilize its filter cavity squeezing source, the frequency-dependence of LIGO's quantum response must be precisely understood.

Degradations to squeezing from optical loss and ``phase noise'' fluctuations of the squeezing angle are also prominently observed in LIGO. Whereas QRPN's correlations cause frequency dependent effects, loss and phase noise are typically described as causing frequency independent, broadband changes to the quantum noise spectrum. This work analyzes the quantum response of both LIGO interferometers to injected squeezed states, indicating that QRPN and broadband degradations, taken independently, are insufficient to fully describe the observed quantum response to squeezing.

The first sections of this work expand the response and degradation model of squeezing to examine and explain the LIGO quantum noise data by decomposing it into independent, frequency-dependent parameters. The latter sections relate the parameter decomposition back to interferometer models, to navigate how squeezing interacts with cavities that have internal losses, transverse-mode selectivity, and radiation pressure interactions. The spectra at LIGO are explained using a set of broadly applicable analytical expressions, without the need for elaborate and specific computer simulations. The analytical models elucidate the physical basis of LIGO's squeezed state degradations, prioritizing transverse-mode quality using wavefront control of external relay optics
\cite{Cao-OEO20-EnhancingDynamic, Cao-AOA20-HighDynamic, Perreca-PRD20-AnalysisVisualization} to further improve quantum noise. This analysis also demonstrates the use of squeezing as a diagnostic tool\cite{Mikhailov-PRA06-NoninvasiveMeasurements}, examining not only the cavities but also the radiation pressure interaction. These diagnostics show further evidence of the benefit of balanced homodyne detection \cite{Fritschel-OEO14-BalancedHomodyne}, another planned component of the ``A+'' upgrade. The description of squeezing in this work expands the modeling of degradations in filter cavities \cite{Kwee-PRD14-DecoherenceDegradation}, explicitly defining an intrinsic, non-statistical, form of dephasing. Finally, the derivations of the quantum response metrics in sec. \ref{sec:derivations} show how to better utilize internal information inside interferometer simulations, simplifying the analysis of squeezing degradations for current and future gravitational wave detectors.

\section{Squeezing Response Metrics}
\label{sec:metrics}

To introduce the frequency-dependent squeezing metrics, it is worthwhile to first describe the metrics used for standard optical squeezing generated from an optical parametric amplifier (OPA), omitting any interferometer. For optical parametric amplifiers, the squeezing level is determined by three parameters. The first is the normalized nonlinear gain, $y$, which sets the squeezing level and scales from 0 for no squeezing to 1 for maximal squeezing at the threshold of amplifier oscillation. For LIGO, $y$ is determined from a calibration measurement of the parametric amplification \cite{Xiao-PRL87-PrecisionMeasurement, Aoki-OEO06-Squeezing946nm, Takeno-OEO07-ObservationDB, Khalaidovski-CQG12-LongtermStable, Schnabel-OC04-SqueezedLight, Dwyer-OEO13-SqueezedQuadrature}. The second parameter is the optical efficiency $\eta$ of states from their generation in the cavity all the way to their observation at readout. Losses that degrade squeezed states are indicated by $\eta < 1$. Finally, there is the squeezing phase angle, $\phi$, which determines the optical field quadrature with reduced noise and the quadrature with the noise increase mandated by Heisenberg uncertainty, anti-squeezing. By correlating the optical quadratures, variations in $\phi$ continuously rotate between squeezing and anti-squeezing. These parameters relate to the observable noise as:
\begin{align}
  N(\phi) &= \left( 1 - \frac{4 \eta y}{(1 + y)^2} \right)\cos^2(\phi) + \left( 1 + \frac{4 \eta y}{(1 - y)^2} \right)\sin^2(\phi)
\end{align}
%
The noise, $N(\phi)$, can be interpreted as the variance of a single homodyne observation of a single squeezed state, but for a continuous timeseries of measurements, $N$ can be considered as a power spectral density, relative to the density of shot-noise. Using relative noise units, $N=1$ corresponds to observing vacuum states rather than squeezing. While the nonlinear gain parameter $y$ may be physically measured and is common in experimental squeezing literature, theoretical work more commonly builds states from the squeezing operator, parameterized by $r$, which constructs an ideal, ``pure'' squeezed state that adjusts the noise power by $e^{\pm 2r}$. State decoherence due to optical efficiency is then incorporated as a separate, secondary process. This is formally related to the previous expression using:
\begin{align}
  N(\phi)
  &= \eta\left(
    e^{-2r}\cos^2(\phi) + e^{+2r}\sin^2(\phi)
    \right) + \left(1 - \eta\right)
    \label{eq:etaSQZ_basic1}
  \\
  e^{-2r}
  &=
    1 - \frac{4y}{(1 + y)^2} \label{eq:OPA_y_SQZ}
    , \hspace{2.5em} e^{+2r} = 1 + \frac{4y}{(1 - y)^2}
\end{align}
In experiments, the squeezing angle drifts due to path length fluctuations and pump noise in the amplifier, but is monitored using additional coherent fields at shifted frequencies and stabilized by feedback control. This stabilization is imperfect, resulting in a root-mean-square (RMS) phase noise,  $\phiRMSsq$, that mixes squeezing and antisqueezing. Using $\hat{\phi}$ to represent the statistical distribution of the squeezing angle, and $E[\cdot]$ the expectation operation, phase noise can be incorporated as a tertiary process given the expectation values:
\begin{align}
 \phiRMSsq &= E\left[\sin^2(\delta \hat{\phi})\right]
             &
 \Ophi &= E\left[\hat{\phi}\right]
  &
\delta \hat{\phi} &= \hat{\phi} - \Ophi
\end{align}
resulting in the ensemble average noise $\overline{N}$, relative shot noise.
\begin{align}
  \overline{N}(\phi) &= E\left[N(\Ophi + \delta \hat{ \phi})\right]
              \\
   &= \eta\left(1 - \phiRMSsq\right)\left(e^{-2r}\cos^2(\Ophi) + e^{+2r}\sin^2(\Ophi) \right)
     \nonumber\\ &\hspace{1em}
                   + \eta\phiRMSsq \left(e^{+2r}\cos^2(\Ophi) + e^{-2r}\sin^2(\Ophi) \right) + (1 - \eta)
    \label{eq:etaSQZ_basic2}
\end{align}
Again, the relative noise $\overline{N}$ is computed as a single value here, but represents a power spectral density that is experimentally measured at many frequencies. These equations, as they are typically used, represent a change to the quantum noise that is constant across all measured frequencies. Notably, the $\phiRMSsq$ phase noise term, which caps at $1/2$, enters as a weighting factor that averages the anti-squeezing noise increase with squeezing noise reduction, while $\eta$ mixes squeezing with standard vacuum.

Incorporating an interferometer such as LIGO requires extending these equations to handle frequency-dependent effects. The equations must include terms to represent multiple sources of loss entering before, during, and after the interferometer, as well as terms for the frequency-dependent scaling of the quantum noise due to QRPN and the interferometer's suspended mechanics. The extension of the metrics is described by the following equations and parameters:
\begin{align}
  N(\Omega)
  &\equiv \Gamma(\Omega) \cdot \Big( \eta(\Omega) S(\Omega) + \Lambda_\IRO(\Omega) \Big) 
    \label{eq:metric_N}
  \\
  S(\Omega)
  &\equiv S_{\!{-}}\cos^2\Big(\sqzang - \theta(\Omega)\Big) + S_{\!{+}} \sin^2\Big(\sqzang - \theta(\Omega)\Big)
    \\
  S_{\!\pm} &\equiv \big(1 - \Xi'(\Omega)\big)e^{\pm2r} + \Xi'(\Omega) e^{\mp2r}
              \label{eq:metric_S}
  \\
  \Lambda_\IRO(\Omega) &\equiv (1 - \eta_\subI)\eta_\subO\eta_\subR + \eta_\subO(1 - \eta_\subR) + (1 - \eta_\subO) / \Gamma
    \label{eq:metric_Lambda}
\end{align}
These metrics are composed of the following variables:
\renewcommand{\descriptionlabel}[1]{\hspace{\labelsep}{#1}:}
\begin{description}[noitemsep, nolistsep]
 \item[$N(\Omega)$] the power spectrum of quantum noise in the readout, relative to the vacuum power spectral density, $\hbar \omega / 2$, of broadband shot noise.
 \item[$\Gamma(\Omega)$] The quantum noise gain of the interferometer optomechanics. While $N(\Omega)$ is relative shot-noise, QRPN causes interferometers without injected squeezing to exceed shot noise at low frequencies, resulting in $\Gamma > 1$. For optical systems with $\Gamma\ne1$, the system cannot be passive, and must apply internal squeezing/antisqueezing to the optical fields.
 \item[$e^{2r}, e^{-2r}$] The ``pure'' injected squeezing and anti-squeezing level, before including any degradations. This level is computed for optical parametric amplifier squeezers using \cref{eq:OPA_y_SQZ}.
 \item[$S_{\!-}, S_{\!+}$] The minimum and maximum relative noise change from squeezing at any squeezing angle, ignoring losses.
 \item[$S(\Omega)$] The potentially observable injected squeezing level, before applying losses or noise gain.
  \item[$\sqzang$] The frequency independent squeezing angle chosen between the source and readout. This is usually stabilized with a co-propagating coherent control field and feedback system.
  \item[$\theta(\Omega)$] the squeezing angle rotation due to the propagation through intervening optical system. In a GW interferometer, this can be due to a combination of cavity dispersion and optomechanical effects. Quantum filter cavities target this term to create frequency dependent squeeze rotation.
  \item[$\eta_\subI(\Omega)$, $\eta_\subO(\Omega)$, $\eta_\subR(\Omega)$] The individually budgeted transmission efficiencies of the squeezed field at input, reflection and output paths of the interferometer. $1 - \eta_{\text{I,R,O}}$ indicates optical power lost in that component.
  \item[$\eta(\Omega)$] The collective transmission efficiency of the squeezed field. This is usually the product of the efficiencies in each path, $\eta = \eta_\subI\eta_\subO\eta_\subR$, but can deviate from this when $\Gamma \ne 1$ and interferometer losses affect both $\Gamma$ and $\eta_\subR$.
  \item[$\Lambda_\IRO(\Omega)$] The total transmission loss over the squeezing path that contaminates injected squeezed states with standard vacuum. When $\Gamma \approx 1$, then $\Lambda_\IRO \approx 1 - \eta$.
  \item[$\Xi'(\Omega)$] This is a squeezing-level dependent decoherence mechanism called dephasing. It
    incorporates both statistical $\phiRMSsq$ phase fluctuations and the fundamental degradation arises from optical losses with unbalanced cavities, denoted $\Xi(\Omega)$. It can also arise from QRPN with structural or viscous mechanical damping. \Cref{sec:effective_dephasing} shows how to incorporate fundamental dephasing  $\Xi(\Omega)$, standard phase uncertainty, $\phiRMSsq$, and cavity tuning fluctuations, $\thetaRMSsq(\Omega)$, into $\Xi'(\Omega)$ to make a total effective dephasing factor. When small, these factors sum to approximate the effective total $\Xi'$
\end{description}

After the data analysis of the next section, these quantum response metrics are derived in \cref{sec:derivations}. These squeezing metrics indicate three principle degradation mechanisms, all frequency-dependent. These are losses, where $\Lambda_\IRO(\Omega)\approx 1 {-} \eta(\Omega) > 0$; Mis-phasing, from $\phi{-}\theta(\Omega)\ne 0$; and de-phasing, $\Xi(\Omega) > 0$.

The interaction of squeezing with quantum radiation pressure noise is described within these terms. Broadband Squeezing naively forces a trade-off between increased measurement precision and increased quantum back-action. When squeezing is applied in the phase quadrature, it results in anti-squeezing of the amplitude quadrature. The amplitude quadrature then pushes the mirrors and increases QRPN; thus, the process of reducing imprecision seemingly increases back-action. In other terms, QRPN causes the interferometer's ``effective'' observed quadrature\footnote{The observed quadrature in this context is with respect to the injected quantum states, be they squeezed or vacuum. It is dependent on the quadrature of the interferometer's homodyne readout, but does not refer specifically to it. The effective observed quadrature also does not refer to the specific quadrature that the interferometer signal is modulated into.} to transition from the phase quadrature at high frequencies to the amplitude quadtrature at low frequencies. In the context of these metrics, the observation quadrature is captured in the derivation of $\theta(\Omega)$. The associated back-action trade-off can be considered a mis-phasing degradation, allowing the SQL to be surpassed using the quantum quadrature correlations introduced by varying the squeezing angle\cite{Yu-N20-QuantumCorrelations}. Frequency dependent squeezing, viewed as a modification of the squeezing source, can be considered as making $\phi(\Omega)$ frequency-dependent, tracking $\theta(\Omega)$. Alternatively, it can be viewed as a modification of the interferometer, to maintain $\theta(\Omega) \approx 0$. While a quantum filter cavity is not explicitly treated in this work, the derivations of \cref{sec:derivations} are setup to be able to include a filter cavity as a modification to the input path of the interferometer.

While mis-phasing can be compensated using quantum filter cavities, the other two degradations are fundamental. For squeezed states, they establish the noise limit:
\begin{align}
  N(\Omega) &\ge \Gamma{\cdot}\left(2\eta\sqrt{\Xi'(1-\Xi')} + \Lambda_\IRO\right),
  &
  e^{-2r} &= \sqrt{\Xi'(\Omega)}
  \label{eq:N_limit}
\end{align}
Setting the squeezing level as $\sqrt{\Xi'}$ solves for the optimal noise given the dephasing. Squeezing is then further degraded from losses, producing the noise limit. Notably, the optimal squeezing is generally frequency-dependent due to $\Xi(\Omega)$, indicating that for typical broadband squeezing sources, this bound cannot always be saturated at all frequencies.

\subsection{Ideal Interferometer Response}
\label{sec:ideal_IFO_metrics}
Before analyzing quantum noise data to utilize the squeezing metrics of \crefrange{eq:metric_N}{eq:metric_Lambda}, it is worthwhile to first review the quantum noise features expected in the LIGO detector noise spectra\cite{Kimble-PRD01-ConversionConventional, Aasi-CQG15-AdvancedLIGO}, under ideal conditions and without accounting for realistic effects present in the interferometer. The derivations later will then extend how the well-established equations below generalize to incorporate increasingly complex interferometer effects, both by extracting features from matrix-valued simulation models, as well as by extracting features from scalar boundary-value equations for cavities.

\autofiguresvgTEX{
  folder=./figures/, 
  file=SQZ_mm_IFO, 
  label={SQZ_mm_IFO},
  caption={
    This simplified diagram of the interferometer layout shows the propagation of the source laser (solid red) and squeezed beam (dashed burgundy). At (a), the squeezed beam is sourced from a parametric amplifier cavity and circulated to the interferometer with a Faraday isolator. At (b), the squeezing field reflects from the interferometer. Depending on the frequency and transverse beam profile, the states partially transit the interferometer cavities, but also partially reflect promptly. The squeezing that enters the interferometer symmetrically is beam split inside the signal recycling cavity, coherently resonates in both arms, and recombines again at the beamsplitter, effectively experiencing the two branches as a single linear coupled cavity. Injected at a different port, the red laser field carries substantial laser power and is symmetrically split to pump the arm cavities. Differential length signals are sourced by modulating the circulating pump field, creating a phase-quadrature field that resonates in the same effective linear cavity as the squeezing. The signal is emitted at (b) follows with the reflected squeezing. The transverse beam profile (mode) of the signal and squeezing is then selected using the output mode cleaning cavity at (c). Ultimately, the signal and noise are read as timeseries in photodetectors at (d). This effect of coherent interference between prompt and cavity-circulated squeezing from this sequence is formulated, measured, and analyzed in the following sections.
},
}

Other than shot noise imprecision, the dominant quantum effect in gravitational wave interferometers arises from radiation pressure noise. In an ideal, on-resonance interferometer, this noise is characterized by the interaction strength $\K(\Omega)$ that correlates amplitude fluctuations entering the interferometer to phase fluctuations that are detected along with the signal. $\K$ is generated from the circulating arm power $P_\Arm$ creating force noise that drives the mechanical susceptibility $\chi(\Omega)$. The susceptibility relates force to displacement on each of the four identical mirrors of mass $m$ in the GW arm cavities. The QRPN effect is enhanced by optical cavity gain $g(\Omega)$ which resonantly enhances quantum fields entering the arm cavities and signal fields leaving them.
\begin{align}
  \mathcal{K}(\Omega)
  &= 16k\frac{P_\Arm}{c} g^2(\Omega)\chi(\Omega),
    &
  g(\Omega) &= \frac{\sqrt{{\gamma_\Arm c}/{L_{\arm}}}}{\gamma_\Arm + i\Omega}
              \label{eq:optical_gain_g}
\end{align}
Here, $k$ is the wavenumber of the interferometer laser and $c$ the speed of light.
The arm cavity gain $g(\Omega)$ is a function of the signal bandwidth $\gamma_\Arm$, derived later, and the interferometer arm length $L_\arm$. Unlike in past works, this expression of $\K(\Omega)$ here is kept complex, holding the phase shift that arises from the interferometer cavity transfer function. The phase of $\K(\Omega)$ is useful for later generalizations. $\K(\Omega)$ adds the amplitude quadtrature noise power to the phase quadrature fluctuations directly reflected from the interferometer, setting the noise gain $\Gamma(\Omega)$
\begin{align}
  \Gamma(\Omega) &= 1 + |\mathcal{K}(\Omega)|^2
  &
  \theta(\Omega) &= \arctan(|\K(\Omega)|)
                   \label{eq:GammaTheta_standard}
\end{align}
The relationship between $\Gamma(\Omega)$ and $\theta(\Omega)$ from $\K(\Omega)$ is stated above as reference, but it will more appropriately handle the complex $\K(\Omega)$ when it is derived later. The value $|\K(\Omega_\sql)| \equiv 1$ defines the crossover frequency $\Omega_\sql$ between noise contributions from shotnoise imprecision and QRPN, corresponding to $\Gamma(\Omega_\sql)= 2$ and $\theta(\Omega_\sql)=-45^\circ$. For the $\chi(\Omega)$ susceptibility of a free test mass, the factor $\K(\Omega)$ can be expressed using only frequency scales.
\begin{align}
  \mathcal{K}(\Omega)
  &= -\frac{\Omega^2_\sql}{\Omega^2}\left( \frac{\gamma_\Arm}{\gamma_\Arm + i\Omega}\right)^2,
  &&\text{given}&
  \chi(\Omega) &\equiv \frac{-1}{m \Omega^2}
    \label{eq:Kchi_standard}
\end{align}

\begin{figure*}
	\centering
	\includegraphics[width=\textwidth]{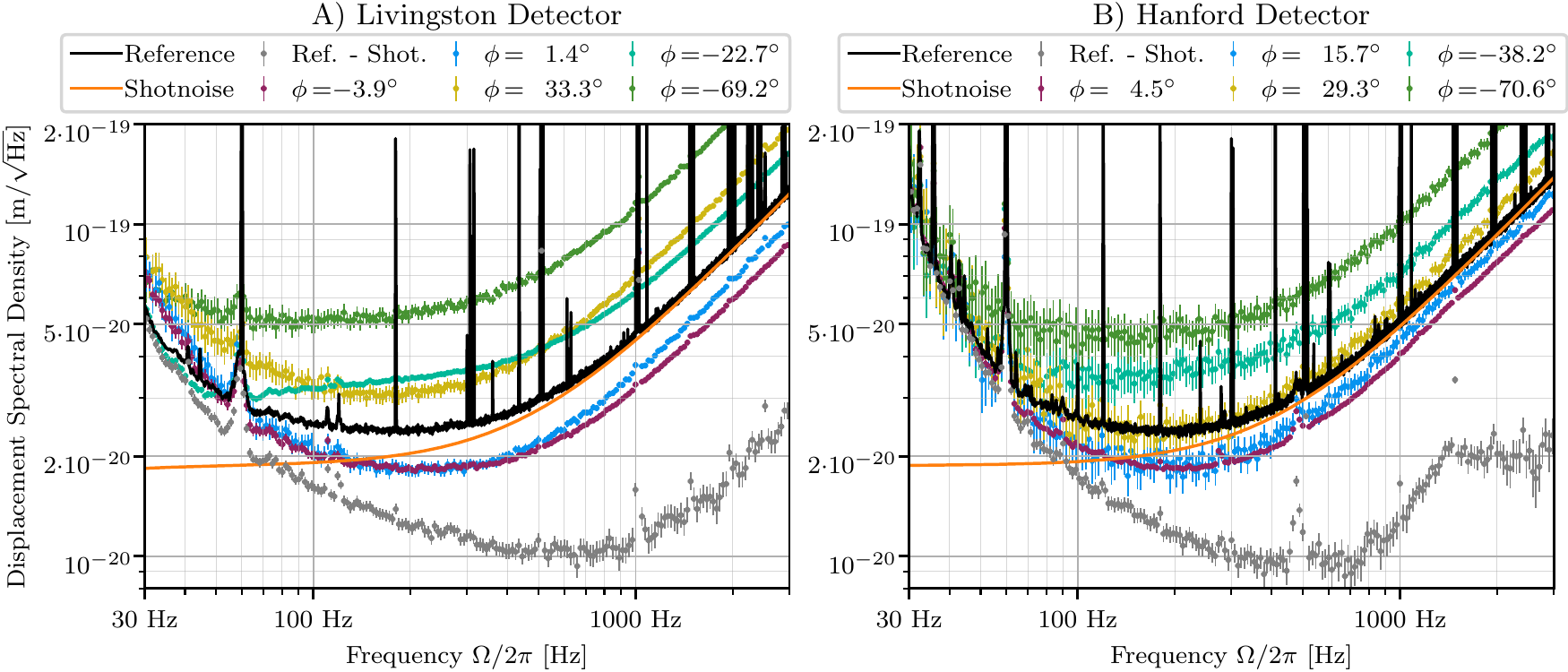}
	\caption{
    This figure plots the total quantum and classical noise measured in the LIGO detectors in displacement amplitude spectral density units. The black trace plots a reference measurement of the total noise without injected squeezing at 0.25Hz resolution over 1.5Hr integration for LLO and 1.1Hr for LHO. The orange shotnoise measurement shows the displacement calibration, $\sqrt{G(\Omega)}$, in amplitude density units. Subtracting the shotnoise level from the reference yields the gray datapoints, which have been rebinned using a median statistic applied after the subtraction and with a logararithmic bin spacing. The subtraction primarily shows the classical noise but also contains QRPN. Multiple measurements are taken at varied squeezing angles, with 5 of 12 plotted for Livingston (LLO) and 5 of 34 plotted for Hanford (LHO), using the same median rebinning method as the gray subtraction. The variation in the data errorbars results from the binning span of each datapoint, $\Delta F$, and the measurement integration time, $\Delta T$. The measured spectra error relative to the total noise and proportional to $1/\sqrt{\Delta F \Delta T}$. The squeezing angle of $-3.9^\circ$ and $1.4^\circ$ datasets at LLO used ${\sim}1$Hr integration, and the remainder used 15 min each. The squeezing angle $4.5^\circ$ dataset at LHO used ${\sim}1$Hr integration, while all others use 2 minutes each. The squeezing level $e^{\pm2r}$ is constant over all angles, but different between the two sites. This accounts for the difference in the yellow, ${\sim}30^{\circ}$, dataset at each site.
  }
	\label{fig:data_h}
\end{figure*}

Frequency independent losses are applied to squeezing before and after the interferometer using $\eta=\eta_\subI\eta_\subR\eta_\subO$ where $\eta_\subI < 1$, $\eta_\subO < 1$. The ideal interferometer assumption of the formulas above enforce $\eta_\subR=1$. Phase noise in squeezing is included in this ideal interferometer case using $\Xi' = \phiRMSsq$.

The above expressions relate the optical noise $N(\Omega)$ of \cref{eq:metric_N} to past models of the quantum strain sensitivity of GW interferometers\cite{Kimble-PRD01-ConversionConventional, Buonanno-PRD01-QuantumNoise, Buonanno-PRD03-ScalingLaw}. Since $N(\Omega)$ is relative to shot-noise, it must then be converted to strain or displacement using the optical cavity gain $g(\Omega)$, by how it affects the GW signal through the calibration factor $G(\Omega)$. This factor $G(\Omega)$ relates strain modulations to optical field phase modulations in units of optical power.
\begin{align}
  \text{PSD}_{\text{strain}}(\Omega) &= G(\Omega)N(\Omega),
                                       &
  G(\Omega) &= \frac{\hbar c}{\eta_\subO L_{\arm}^2|g(\Omega)|^{2} k P_\Arm }
              \label{eq:optical_calibration_G}
\end{align}
Together, these relations allow one to succinctly calculate the effect of squeezing on the strain power spectrum in the case of an ideal interferometer. These factors and the calculations behind them will be revisited as non-idealities are introduced.

\section{Experimental Analysis and Results}
\label{sec:experiment}

A goal of this paper is to use the squeezing response metrics of \crefrange{eq:metric_N}{eq:metric_Lambda} to relate measurements of the instrument's noise spectrum to the parameters of the squeezer system, namely its degradations due to loss $1-\eta$, radiation pressure from mis-phasing $\phi{-}\theta(\Omega)$, and dephasings $\Xi'(\Omega)$. This section presents measurements from the LIGO interferometers that are best described using the established frequency-dependent metrics. The measurements then motivate the remaining discussion of the paper that construct simple interferometer models to describe this data in the context of the metrics. This section refers to and relates to the later sections to provide early experimental motivation for the discussions that follow. The reader may prefer instead to skip this section and first understand the models before returning to see their application to experimental data.

The main complexity in analyzing the LIGO data is that the detectors have additional classical noises, preventing a direct measurement of $N(\Omega)$. The many frequency-dependent squeezing parameters must also be appropriately disentangled. To address both of these issues, the unknown squeezing parameters are fit simultaneously across multiple squeezing measurements. The classical noise contribution is determined by taking a reference dataset where the squeezer is disabled, such that $S(\Omega)=1$, and then subtracting it from the datasets where squeezing is injected.

\begin{figure*}
	\centering
	\includegraphics[width=\textwidth]{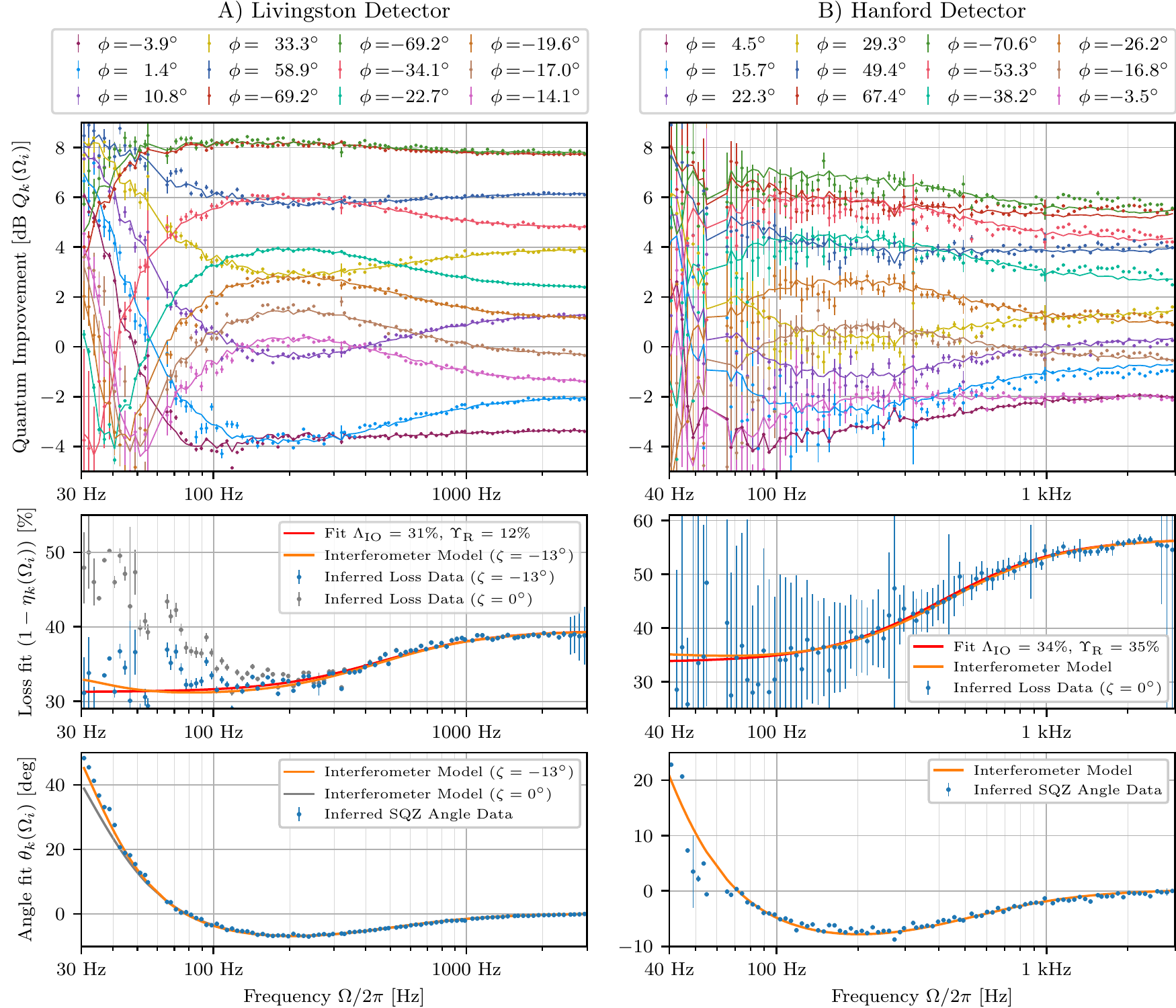}
  \caption{
    This figure shows the data of \cref{fig:data_h} processed as per \cref{sec:experiment} for each LIGO site. The processing subtracts away the classical noise determined from the unsqueezed reference dataset. The top panels show the relative noise change $Q_k(\Omega_i)$ of \cref{eq:Q_processing} computed using $\Gamma_k(\Omega_i)$ from the exact interferometer model of \cref{sec:matrix_coupled_cavity} using the parameters of \cref{tab:LIGO_params}. The top panel includes dots with errorbars for the processed data and lines for the best-fit $Q_k(\Omega_i)$. The middle panel shows the best-fit frequency dependent loss as data points, with errorbars propagated through the fit. For LLO, two sets of loss datapoints are shown, corresponding to interferometer models with different readout angles $\zeta$.
    The loss plots also show $1{-}\eta(\Omega)$ as computed from the exact matrix model, along with a phenomenological fit against the model of \cref{eq:Lambda_MM} of \cref{sec:modeling_TMM}. The phenomenological fit assumes frequency independent losses from the input and output squeezing path with a frequency-dependent addition attributed to transverse mismatch. The bottom panels show the frequency-dependent fit to the observed squeezing angle $\theta_k(\Omega_i)$, using the convention of $\theta(2\pi{\cdot}3\text{kHz}) = 0$. It also plots $\theta(\Omega)$ as computed using the exact matrix model. For the LLO data, the $\zeta=0^\circ$ model is typically assumed for Michelson-like interferometers such as LIGO; However, the model at that readout angle implies losses at low frequencies that are not favored by the $\eta(\Omega)$ models explored in this paper. Alternatively, the $\zeta\approx -13^\circ$ model is consistent with both the fitted losses and the fitted squeezing angles.
  }
  \label{fig:data_Q}
\end{figure*}

Representative strain spectra from the LIGO Livingston (LLO) and LIGO Hanford (LHO) observatory datasets are plotted in \cref{fig:data_h}. The Livingston dataset is also reported in \cite{Yu-N20-QuantumCorrelations}, which details the assumptions and error propagation for the classical noise components and calibration. Only statistical uncertainty is considered in this analysis, in order to propagate error to the parameter fits. The strain spectra of \cref{fig:data_h} include a reference dataset where the squeezer is disabled, shown in black and at the highest frequency resolution. Additionally, the shotnoise ($N=1$) is plotted in orange, indicating the calibration $\sqrt{G(\Omega)}$ of \cref{eq:optical_calibration_G}. The gray subtraction curve depicts the total classical noise contribution summed with the radiation pressure noise $G(\Omega)\K^2(\Omega)$. The gray dataset can equivalently be computed using a cross correlation of the two physical photodetectors at the interferometer readout\cite{martynov-pra17-quantumcorrelation}. The equivalence of subtraction and cross correlation is used to precisely experimentally determine the shot-noise scale $G(\Omega)$ from the displacement-calibrated data.

\subsection{Analysis}

Each squeezing measurement, indexed by $k$, is indicated by $M_{\text{ref}, k}(\Omega_i)$, with a value at each frequency indexed by $i$. The reference dataset is denoted $M_{\text{ref}}(\Omega_i)$. The two are subtracted to cancel the stationary classical noise component. The calibration $G(\Omega)$ is removed to result in the differential quantum noise measurement $D_k(\Omega_i)$.
\begin{align}
  D_k(\Omega_i)
  &\equiv
    \frac{M_{\text{sqz}, k}(\Omega_i) - M_{\text{ref}}(\Omega_i)}{G(\Omega_i)}
    \label{eq:D_processing}
\end{align}
For these datasets, the squeezing level $e^{\pm 2r}$, is held constant and independently measured using the nonlinear gain technique\cite{Dwyer-OEO13-SqueezedQuadrature} to derive $y$ of \cref{eq:OPA_y_SQZ}. Each differential data $D_k(\Omega_i)$ is taken at some squeezing angle $\phi_k$, which is either fit (LLO) or derived from independent measurements (LHO). The parameters $\eta_i$ and squeezing rotation $\theta_i$ are independent at every frequency $\Omega_i$ but fit simultaneously. All $\phi_k$ are also fit simultaneously across all datasets. Nonlinear least squares fitting was performed using the Nelder-Mead simplex algorithm \cite{Gao-COA12-ImplementingNelderMead} implemented in SciPy \cite{Virtanen-NM20-SciPyFundamental}. The residual minimized by least squares fitting is
\begin{align}
  \mathcal{R} &= \sum_{\substack{i=0\\k=0}}^{\mathcal{N}} \left( \frac{D_k(\Omega_i) - \overline{D}_k(\Omega_i)}{\Delta D_k(\Omega_i)}\right)^2
\end{align}
The measurement statistical uncertainty $\Delta D$, dominated by the statistical uncertainty in power-spectrum estimation, was propagated through the datasets per \cite{Yu-N20-QuantumCorrelations}. $\overline{D}_k(\Omega_i)$ is the model of the data that is a function of the fit parameters, $\eta_i, \theta_i, \phi_k$ as well as independently measured parameters such as $e^{-2r}$. $\Xi'(\Omega_i)$ is not fit using this data since the squeezing level $e^{2r}$ is not varied across the datasets. This is discussed below. These given fit parameters affect are propagated through the squeezing metric functions create a model of this particular differential quantum noise measurement.
\begin{align}
  \overline{S}_k(\Omega_i) &\equiv
  e^{-2r}\cos^2\Big(\sqzang_k - \theta_i\Big) + e^{+2r}\sin^2\Big(\sqzang_k - \theta_i\Big)
  \\
  \overline{D}_k(\Omega_i)
  &\equiv \left.N(\Omega_i)\right|_{S = \overline{S}_k(\Omega_i)} - \left.N(\Omega_i)\right|_{S = 1}
\end{align}
Which simplifies to
\begin{align}
  \overline{D}_k(\Omega_i)
  &= \left(\overline{S}_k(\Omega_i) - 1\right) \eta_i \Gamma(\Omega_i)
\end{align}
Notably, the individual efficiencies $\eta_\subI, \eta_\subR, \eta_\subO$ cannot be individually measured and only the ``total'' efficiency $\eta(\Omega)$ is measurable using this differential method, where the classical noise is subtracted using a reference dataset with squeezing disabled. Additionally, the optical efficiency $\eta$ can only be inferred given some knowledge or assumption of $\Gamma(\Omega)$. In effect, the product $\eta \Gamma$ is the primary measurable quantity, rather than its decomposition into separate $\eta$ and $\Gamma$ terms; However, for the purposes of modeling, decomposing the two is conceptually useful. Furthermore, to characterize physical losses, the efficiency $\eta$ or loss $\Lambda_\IRO\approx 1 {-} \eta$ is easier to plot and interpret than the product $\eta\Gamma$.

For these reasons, the differential data $D_k(\Omega_i)$ is further processed, creating the measurement $Q_k(\Omega_i)$ with a form similar to \cref{eq:etaSQZ_basic1}
\begin{align}
  Q_k(\Omega_i)
  &\equiv 
    \frac{D_k}{\Gamma(\Omega_i)} + 1 \approx S_k(\Omega_i)\eta_i + (1 - \eta_i) + \Delta Q
    \label{eq:Q_processing}
\end{align}
The LIGO squeezing data expressed in dB's of $Q_k(\Omega_i)$ are plotted in the upper panels of \cref{fig:data_Q}. The data and error bars are in discrete points, while the parameter fits to $Q_k$ using $\eta_i$, $\theta_i$ and $\phi_k$ are the solid lines between the data points. The spectra in each set are calculated using the Welch method a median statistic at each frequency to average all of the frames through the integration time. This prevents biases due to instrumental glitches adding non-stationary classical noise. This technique is detailed in \cite{Yu-N20-QuantumCorrelations}.

After computing $Q_k(\Omega_i)$ at full frequency resolution, the data is further rebinned to have logararithmic spacing by taking a median of the data points within the frequency range of each bin. This rebinning greatly improves the statistical uncertainty at high frequencies, where many points are collected. At lower frequencies, the relative error benefits less from binning; however, both the LLO and the LHO datasets use a long integration time for their reference measurement and at least one of the squeezing angle measurements. Using the median removes narrow-band lines visible in the strain spectra of \cref{fig:data_h}. Fitting combines the few long-integration, low-error datasets with many short-integration, high-error sets at many variations of the operating parameters. The few low-error datasets reduce the absolute uncertainty in the resulting fit parameters, whereas the many variations reduce co-varying error that would otherwise result from modeling parameter degeneracies.

The relative statistical error in each bin of the original PSD $M_k(\Omega_i)$ is approximately $(\Delta F \Delta T)^{-{1}/{2}}$ given the integration time $\Delta T$ of 2 minutes to 1 hour and bin-width $\Delta F$ of 0.25Hz. This relative error is converted to absolute error and propagated through the processing steps of \crefrange{eq:D_processing}{eq:Q_processing}. At low frequencies, the classical noise contribution to each $M_k$ is larger than the quantum noise. Although it is subtracted away to create $D_k(\Omega_i)$, the classical noise increases the absolute error, and, along with less rebinning, results in the larger relative errors at low-frequency in \cref{fig:data_Q}. After fitting the squeezing parameters, the Hessian of the reduced chi-square is computed from the Jacobian of the fit residuals with respect to the parameters. This Hessian represents the Fisher information, and the diagonals of its inverse provides the variances indicated by the plotted loss and angle parameter error bars.

For the LHO data, the fit parameters $\phi_k$ are determined by mapping the demodulation angles of its coherent control feedback system\cite{Tse-PRL19-QuantumEnhancedAdvanced, Dooley-OEO15-PhaseControl, Vahlbruch-PRL06-CoherentControl, Chelkowski-PRA07-CoherentControl} back to the squeezing angle. That mapping has 3 unknown parameters, an offset in demodulation angle, an offset in squeezing angle, and a nonlinear compression parameter, all of which are fit simultaneously in all datasets. This $\phi_k$ mapping was not performed on the LLO data, as some systematic errors in the demodulation angle records bias the results. Despite fitting more independent parameters, the longer integration time of the LLO data gives it sufficiently low statistical uncertainty at frequencies below $\Omega_\sql$ that the model and parameter degeneracy between $\phi_k$, $\theta_i$ and $\eta_i$ is not an issue.

\subsection{Results}

The middle panels of \cref{fig:data_Q} show the fits to $\eta_i$, though plotted as loss $1{-}\eta_i$ to represent $\Lambda_\IRO$. Both datasets additionally include a red loss model curve fit, assembled using the equations in \cref{sec:modeling}. The orange exact model curves use \cref{sec:matrix_coupled_cavity}. The data and model curve fit shows a variation in the efficiency, where losses increase from low to high frequencies. This increase in loss can be attributed either to losses within the signal recyling cavity of the interferometer, or to a coherent effect resulting from transverse Gaussian beam parameter mismatch between the squeezer and interferometer cavities. At low frequencies, the optical efficiency is similar between the two LIGO sites, indicating that frequency independent component to the loss are consistent between the implementations at both LIGO sites. The differing high-frequency losses can reasonably be ascribed to variations in the optical beam telescopes of the squeezing system and are analyzed in \cref{sec:modeling_TMM}.

The LLO middle panel of \cref{fig:data_Q} shows two separate inferred loss $1 - \eta_i$ datasets. These differ in their underlying model of $\Gamma(\Omega_i)$. The following section \ref{sec:derivations} discusses how variations in $\Gamma$ arise and describes the local oscillator angle $\zeta$. The $\zeta = 0$ data reflects the standard, ideal radiation pressure noise model of \crefrange{eq:Kchi_standard}{eq:GammaTheta_standard}. This model is disfavored given the frequency dependency of $\eta(\Omega_i)$ derived using optical cavity models later in this paper. The $\zeta = -13^\circ$ model presents an alternative that is compatible with models of the optical efficiency. The need for this alternative indicates that squeezing metrics must account for variations in interferometer noise gain $\Gamma$. Physically, these variations arise from the readout angle adjusting the prevalence of radiation pressure versus pondermotive squeezing. The $\zeta = -13^\circ$ model results in a smaller noise gain $\Gamma$ at $40 \text{Hz}$ than does the $\zeta = 0^\circ$ model. Since the lower $\Gamma$ model is favored, this dataset provides some, moderate, evidence that LLO currently benefits from the quantum correlations introduced by the mirrors near $\Omega_{\text{SQL}}$, while experiencing lessened sensitivity elsewhere.

This data demonstrates that the readout angle has an effect on the interferometer sensitivity and the optimal local oscillator is not necessarily $\xi = 0$ due to radiation pressure.  The quantum benefit of decreased $\Gamma$ from the readout angle $\xi$ is a method to achieve sub-SQL performance that is an alternative to injecting squeezing. Like squeezing, it has a frequency dependent enhancement known as the ``variational readout'' technique \cite{Kimble-PRD01-ConversionConventional, Khalili-PRD07-QuantumVariational}, that a sensitivity increase from lowering $\Gamma$ while minimizing the sensitivity decrease of the frequency-independent form. For LLO, the reduced sensitivity from $\xi \ne 0$ masquerades as a $5\%$ loss of signal power, but does not actually affect the $\eta$ or $\Lambda_\IRO$ contributions to the squeezing level.

The bottom panels of \cref{fig:data_Q} show the fits of $\theta_i$ of each dataset. The magnitude of $e^{\pm 2r}$ provides a ``lever arm'' in the variation of $S_k(\Omega_i)$ that strongly constrains the $\phi_k{-}\theta_i$ effective squeezing angle. These leveraged constraints result in small errorbars to the fitted $\theta_i$. The LLO data are plotted with two models of the $\theta_i$ based on the assumed local oscillator angle $\zeta$. The $\zeta=0^\circ$ model follows the standard radiation pressure model of \cref{eq:GammaTheta_standard} at low frequencies and includes a filter-cavity type rotation around the interferometer cavity bandwidth $\gamma \approx 2\pi \cdot 450\text{ Hz}$. This rotation is modeled in \cref{sec:modeling_cavities}. The $\zeta=-13^{\circ}$ model is computed using the coupled cavity model of \cref{sec:matrix_coupled_cavity} and internally includes a weak optical spring effect along with the shifted readout angle $\zeta$. Together, these effects modify the effective squeezing angle $\theta$ away from \cref{eq:GammaTheta_standard} at low frequencies, and agree well with the dataset. This agreement provides further evidence of the reduced radiation pressure noise gain $\Gamma(\Omega)$ in LLO that results from the effective LO readout angle $\zeta$. A nonzero readout angle $\zeta$ is reasonable to expect due to unequal optical losses in the LIGO interferometer arm cavities. The arm mismatch results in imperfect subtraction of the fringe-light amplitude quadrature at the beamsplitter, creating a static field that adds to the phase-quadrature light created from the Michelson offset and results in $\zeta \ne 0$. Past diagnostic measurements
conclude that some power in the readout diodes must be in the amplitude quadrature, but until now could not determine the sign.

Although the squeezing angle parameters $\phi_k$ and $\theta_i$ are fit, the frequency-dependent dephasing parameter $\Xi_i$ cannot be reliably determined from these datasets given the accuracy to which $e^{\pm 2r}$ is measured. Additionally, the squeezing level $e^{\pm 2r}$ is not varied in this data, nor is it sufficiently large to resolve an influence from $\Xi(\Omega) < 10^{-3}$. This $\sqrt{\Xi} \approx \phiRMS$ is expected from independent measurements of phase jitter that propagate through the coherent control scheme of the squeezer system\cite{Tse-PRL19-QuantumEnhancedAdvanced}. A large source of optically induced $\Xi$ is not expected has the interferometer cavities are not sufficiently detuned. Measurements of the squeezing system indicate $\phiRMS \lesssim 30\text{ mRad}$. Future LIGO measurements should include additional datasets that vary $r$ along a third indexing axis $j$ and should increase the injected squeezing level $e^{2r} > 30$ to measure, or at least bound, $\Xi$ and its frequency-independent contribution $\phiRMSsq$. The model fits described above are consistent with the data while assuming $\Xi = \phiRMSsq \equiv 0$.

\section{Decomposition Derivation}
\label{sec:derivations}

The factors $\eta(\Omega), \theta(\Omega)$ and $\Xi(\Omega)$ from \cref{sec:metrics} each describe an independent way for squeezing to degrade. $\Gamma(\Omega)$ indicates how the quantum noise scales above or below the shot noise level from squeezing and from quantum radiation pressure within the interferometer. They represent a natural extension of standard squeezing metrics that incorporates frequency dependence, and, as scalar functions, they are simple to plot and to relate with experimental measurements. This section delves into their derivation by employing matrices in the two photon formalism \cite{Caves-PRA85-NewFormalism, Schumaker-PRA85-NewFormalism} to represent the operations of squeezing, adding loss, shifting the squeezing phase, reflecting from the interferometer, and final projection of the quantum state into the interferometer readout. The derived formulas can be used in frequency-domain simulation tools that compute noise spectra using matrix methods, so that the quantum response metrics can be provided in addition to opaquely propagating squeezing to an simulation result of $N(\Omega)$. 

\autofiguresvgTEX{
  folder=./figures/, 
  file=SQZ_mm_cavities, 
  label=SQZ_mm_cavities, 
  caption={
    The two-photon transformation matrices experienced by squeezing through the sequence of \cref{fig:SQZ_mm_IFO}. The effective linear coupled cavity, including the optomechanical effect of radiation pressure, is collected and computed into the transformation $\Tmat{H}_R$. The middle cavity is the signal recycling cavity and the rightmost cavity represents the coherent combination of both arms. Each cavity adds losses from each mirror. For simplicity, these are collected into round-trip cavity loss contributions, $\Lambda_{\subR, \src}$, and $\Lambda_{\subR, \src}$ that inject standard optical vacuum into the cavities, circulating and transforming into the loss terms $\Tmat{T}_{\subR,\smu}$ while lowering the efficiency $\eta_\subR$. Transformations of the squeezing at the input and output are included with the terms, $\Tmat{H}_{\subI}$, $\Tmat{H}_{\subO}$ and any additive vacuum contributions,  $\Tmat{T}_{\subI,\smu}$, $\Tmat{T}_{\subO,\smu}$.
  },
}

Two-photon matrices are an established method to represent transformations of the optical phase space of Guassian states in an input-output Heisenberg representation of the instrument\cite{Danilishin-LRR12-QuantumMeasurement}. They are concise yet rigorous when measuring noise spectra from squeezed states using the quantum measurement process of Homodyne readout. Section II of \cite{Danilishin-LRR19-AdvancedQuantum} provides a review of their usage in the context of gravitational-wave interferometers. Here, two-photon matrices are indicated by doublestruck-bold lettering, and are given strictly in the amplitude/phase quadrature basis.

Each matrix represents the transformation of the optical phase space of a single optical ``mode'' as it propagates through each physical element towards the readout. The term ``mode'' refers to a basis vector in a linear decomposition of optical field the transverse optical plane of many physical ports\footnote{Simulations also typically span multiple optical frequencies, but this is not treated here.}. Each plane is further decomposed into transverse spatial modes using a Hermite or Laguerre-Gaussian basis. In this decomposition, each optical mode is indexed by the placeholder $\mu$ and acts as a continuous transmission channel for optical quantum states. The phase space transformations of these continuous optical states is indexed by time or, more conveniently, frequency. Optical losses and mixing from transverse mismatch behave like beamsplitter operations, serving to couple multiple input modes, generally carrying vacuum states, to the mode of the readout where states are measured.

The mode of the injected squeezed states, and their specific transformations during beam propagation, must be distinguished from all of the lossy elements that couple in vacuum states. The squeezed states experience a sequence of transformations by the input elements, interferometer, and output elements, denoted $\Tmat{H}_\subI(\Omega)$,  $\Tmat{H}_\subR(\Omega)$,  $\Tmat{H}_\subO(\Omega)$. This sequence multiplies to formulate the total squeeze path propagation $\Tmat{H}$. 
\begin{align}
  \Tmat{H}(\Omega) &= 
\Tmat{H}_{\text{O}}\Tmat{H}_{\text{R}}\Tmat{H}_{\text{I}}
\end{align}
Lossy optical paths mix the squeezed states with additional standard vacuum states. These are collected into sets of transformation matrices corresponding to each individual loss source, $\{\Tmat{T}_\mu\}$. See \cref{fig:SQZ_mm_cavities}. The sets are grouped by their location along the squeezing path where the lossy element is incorporated. The beamsplitter-like operation that couples each loss is given by a $\Tmat{\Lambda}_\mu$, indexed by its location and source along the squeezing path. Loss transformations $\Tmat{\Lambda}_\mu$ are generally frequency-independent. $\Tmat{\Lambda}_{\text{R}, i}$ are an exception, as they occur within the cavities of the interferometer and include some cavity response. The vacuum states associated with each loss then propagate along with the squeezed states and experience the remaining transformations that act on squeezing.
\begin{align}
  \Tmat{T}_{\text{I},\smu}(\Omega)
  &= 
\Tmat{H}_{\text{O}}\Tmat{H}_{\text{R}}\Tmat{\Lambda}_{\text{I},\smu}
    \\
  \Tmat{T}_{\text{R},\smu}(\Omega)
  &= 
\Tmat{H}_{\text{O}}\Tmat{\Lambda}_{\text{R},\smu}
    \\
  \Tmat{T}_{\text{O},\smu}(\Omega)
  &= 
\Tmat{\Lambda}_{\text{O},\smu}
  \\
  \left\{\Tmat{T}\right\} &= \left\{\Tmat{T}_{\text{I},\smu}; \Tmat{T}_{\text{R},\smu}; \Tmat{T}_{\text{O},\smu}  \right\}
\end{align}
Together, all of the transformations of $\Tmat{H}$ and  $\{\Tmat{T}\}$ define the output states at the readout of the interferometer in terms of the input states entering through the squeezer and loss elements. The two quadrature observables of the optical states are given with the convention $\hat{q}$ being the amplitude quadrature and $\hat{p}$ being the phase, and they are indexed to distinguish their input port and transverse mode.
\begin{align}
  \begin{bmatrix}\hat{q}_{\text{out}}(\Omega) \\ \hat{p}_{\text{out}}(\Omega)\end{bmatrix}
  &=
    \Tmat{H} \begin{bmatrix}\hat{q}_{\text{in}}(\Omega) \\ \hat{p}_{\text{in}}(\Omega)\end{bmatrix} + 
  \sum_{\Tmat{T}_\smu\in \left\{\Tmat{T}\right\}} \Tmat{T}_\smu \begin{bmatrix}\hat{q}_\smu(\Omega) \\ \hat{p}_\smu(\Omega)\end{bmatrix}
  \label{eq:qp_out}
\end{align}
The two-photon matrices $\Tmat{H}$ and $\Tmat{T}_\smu$ must preserve commutation relations, namely $[\hat{q}_\text{out}, \hat{p}_\text{out}] = [\hat{q}_\smu, \hat{p}_\smu] = i\hbar$. In doing so, the matrices ensure that losses within $\Tmat{H}$ couple ancillary vacuum states that degrade squeezing.

The readout carries a continuous coherent optical field known as the ``local oscillator'' and the output states are read using homodyne readout. The phase of the local oscillator, $\zeta$, defines the observed quadrature, $\hat{m}$, for the homodyne measurement. Gravitational Wave interferometers typically use a ``Michelson offset''\cite{Fricke-CQG12-DCReadout, Hild-CQG09-DCreadoutSignalrecycled, Ward-CQG08-DcReadout} in the paths adjacent their beamsplitter to operate slightly off of dark fringe. This offset couples a small portion of their pump carrier light to their output as the local oscillator field. This is a form of homodyne readout that fixes $\zeta$ to measure in the phase quadrature, defined here to be when $\zeta=0$. Imperfect interference at the beamsplitter can couple some amplitude quadrature and shift $\zeta$ away from $0$. Balanced homodyne readout is an alternative implementation proposed for LIGO's ``A+'' upgrade and will allow $\zeta$ to be freely chosen\cite{Fritschel-OEO14-BalancedHomodyne}. Regardless of the implementation, the homodyne observable is $\hat{m}$,
\begin{align}
  \hat{m} &=
  \TLO\begin{bmatrix}\hat{q}_{\text{out}}(\Omega) \\ \hat{p}_{\text{out}}(\Omega)\end{bmatrix}
  &
  \TLO(\zeta)
  &=
    \begin{bmatrix}
      \sin(\zeta) & \cos(\zeta)
    \end{bmatrix}
 \label{eq:homodyne_observable}
\end{align}
Homodyne readout enforces a symmetrized expectation operator, denoted here with the subscript HR, for all measurements of the optical quantum states.
Further details of the measurement process are beyond the scope of this work, but the following quadratic expectations arise when computing the noise spectrum and are sufficient to simplify the homodyne expectation values of $\hat{m}$.
\begin{align}
  1 &= \braket{\hat{q}_\smu^2}_{\text{HR}} = \braket{\hat{p}_\smu^2}_{\text{HR}},
      \label{eq:vacuum_expectations}
  \hspace{1em}
  0 = \braket{\hat{q}_\smu\hat{p}_\smu}_{\text{HR}} = \braket{\hat{p}_\smu\hat{q}_\smu}_{\text{HR}}
      \\
  0 &= \braket{\hat{q}_\smu\hat{q}_\snu}_{\text{HR}} = \braket{\hat{p}_\smu\hat{p}_\snu}_{\text{HR}} \text{  for } \nu \ne \mu
      \label{eq:vacuum_expectations2}
\end{align}
As a result of these expectations, the vector norm suffices to evaluate noise power using this matrix formalism.
The addition of squeezing can be seen either as a modification of the input states $\hat{q}_{\text{in}}$, $\hat{q}_{\text{in}}$, which violate \cref{eq:vacuum_expectations}. This work uses the alternative picture, where an additional squeezing transformation occurring at the very start of the squeezing path $\Tmat{H}$ that acts on $\hat{q}_{\text{in}}$, $\hat{q}_{\text{in}}$ that are also vacuum states. The squeezing transformation is defined by the squeezing level $r$ and the squeezing angle $\phi$, which act via the matrices:
\begin{align}
  \Tmat{R}(\phi)
  &\equiv
  \begin{bmatrix}
    \cos(\phi) & {-}\sin(\phi)\\
    \sin(\phi) & \phantom{-}\cos(\phi)
  \end{bmatrix}
 &
  \Tmat{S}(r) &\equiv
  \begin{bmatrix}
    e^r & 0\\
    0 & e^{-r}
  \end{bmatrix}
\end{align}
When added to the squeezing path, the resulting quantum noise is calculated from the observable $\hat{m}$.
\begin{align}
  N(\Omega) &= \Braket{\hat{m}^\dagger\hat{m}}_{\text{HR}} = 
              \left| \TLO \Tmat{H}\Tmat{R}(\phi) \Tmat{S}(r) \right|^2 + 
\sum_{\Tmat{T}_\smu\in \left\{\Tmat{T}\right\}} \left| \TLO \Tmat{T}_\smu \right|^2
                             \label{eq:N_no_sqz_derivation}
\end{align}
The first term of which is one of the factors in \cref{eq:metric_N}
\begin{align}
  \eta(\Omega)\cdot S(\Omega, \phi) \cdot \Gamma(\Omega, \zeta)
   &= \left|\TLO
     \Tmat{H}\Tmat{R}(\phi)
     \Tmat{S}(r)
     \right|^2
    \label{eq:decomposition_relation}
\end{align}
At this point, the factors can be separated because: $\Tmat{R}\Tmat{S}$ determines the factor $S(\Omega, \phi)$; $\Tmat{H}$ has been ``reduced'' by loss, indicating when $\eta(\Omega) < 1$; and the benchmark noise level is defined by $\Gamma(\Omega)$, contained in the interferometer's optomechanical element $\Tmat{H}_{\text{R}}$.

To distinguish these terms, further manipulations are necessary. The first is to examine just the vector $\TLO \Tmat{H}$ to determine how the later term $\Tmat{R}\Tmat{S}$ results in $S(\Omega)$. Basis vectors for the two quadrature observables are defined, and the local oscillator is represented using them.
\begin{align}
  \TLO(\zeta)
  &=
  \Tvec{e}_p^\dagger\Tmat{R}(\zeta)
  &
    \Tvec{e}_q &= 
    \begin{bmatrix}
      1 \\ 0
    \end{bmatrix}
  &
    \Tvec{e}_p &= 
    \begin{bmatrix}
      0 \\ 1
    \end{bmatrix}
\end{align}
The basis vectors then allow the vector norm to be split into its two components $\Mq$ and $\Mp$, defining the \textit{observed noise quadrature}.
\begin{align}
  \Mq(\Omega) &= \TLO \Tmat{H} \Tvec{e}_q & \Mp(\Omega) &= \TLO \Tmat{H} \Tvec{e}_p
\end{align}
The vector $\vec{m}$ contains the magnitude and angle of a projection of the quantum state $\hat{q}_{\text{in}}$, $\hat{p}_{\text{in}}$ at each frequency, but it also contains the complex phase shift from propagation delay in the interferometer and squeezing path. This later phase contribution does not affect noise calculations, but must be properly handled. Projecting it away requires maintaining phase information, and this is why the optomechanical factor $\K$ is complex in this work.

The squeezing angle rotation $\Tmat{R}(\phi)$ can be viewed through its left-multiplication, applying a rotation to the observed noise quadtrature rather than to the squeezing. In this picture, the angle $\phi$ can align the observed quadtrature with either the squeezing or anti-squeezing quadrature. The rotation needed to do so determines $\theta(\Omega)$, again with the caveat that both $m_q$ and $m_p$ are complex. Their common phase carries the delay information, but their differential phase causes dephasing. In short, differential phase forces $\vec{m}$ to project into both quadratures at any rotation $\Tmat{R}(\phi)$. This has the effect of always adding anti-squeezing to squeezing and vice-versa, resulting in the factor $\Xi(\Omega)$. The relations are fully derived in \cref{sec:phase_noise_composition} using a singular value decomposition to identify the principle noise axes. It leads to the expressions
\begin{align}
  \theta(\Omega) &\approx -\arctan\left(\Re\!\left\{{\frac{\Mq}{\Mp}}\right\}\right)
                   \label{eq:theta_calculation}
  \\
  \Xi(\Omega) &= \frac{1}{2} -  \sqrt{
        \frac{\left(|\Mq|^2 - |\Mp|^2\right)^2 + 4 \Re\left\{ \Mq\cMp \right\}}
        {4\left(|\Mq|^2 + |\Mp|^2\right)^2}
        } 
                   \label{eq:Xi_calculation}
\end{align}
The observation vector $\vec{m}$, and \cref{eq:theta_calculation} generalizes the observed noise quadrature description of quantum radiation pressure noise. With it, the observed quadrature angle $\theta(\Omega)$ may be computed for any readout angle $\zeta$ and for more complex interferometers $\Tmat{H}_\subR$. The ideal interferometer example is demonstrated in \cref{sec:ideal_IFO_example}

The phase and magnitude of of the previous argument allows one to determine $S(\Omega)$ from the form of $\Tmat{S}$ applied to $\vec{m}\Tmat{R}(\phi)$. Factoring $S$ away, the magnitude of $\vec{m}$ carries the efficiency of transmitting the squeezed state, along with the noise gain applied to it.
\begin{align}
  \eta(\Omega) \cdot \Gamma(\Omega, \zeta)
  &=
    |\Mq|^2 + |\Mp|^2
    \label{eq:eta_gamma}
\end{align}
$\Gamma(\Omega)$ expresses the total noise from the interferometer when squeezing is not applied, applying radiation pressure or optomechanical squeezing to both the squeezing path vacuum and internally loss-sourced vacuum. $\eta \Gamma$ is affected by all losses, but some of them affect $\Gamma(\Omega)$ as well. Using squeezing or a coherent field to probe $\Tmat{H}$ always measures the product $\eta \Gamma$, so the noise gain factor $\Gamma$ serves primarily as a benchmark. As a benchmark, it relates the dependence of $N(\Omega)$ to $S(\Omega)$ and separates the scaling by the efficiency $\eta$ so that the physical losses may be determined. For this reason, there is freedom to define $\Gamma$ to make it as independent from the losses as possible, so that it best serves as a benchmark. Here, it is defined using the simulated knowledge of the total noise from the interferometer elements alone:
\begin{align}
 \Gamma(\Omega) 
  &= 
\left|\TLO \Tmat{H}_{\text{R}}\right|^2 + 
  \sum_{i}\left|\TLO \Tmat{\Lambda}_{\text{R},\smu}\right|^2
    \label{eq:gamma_HR}
\end{align}
$\eta$ is then determined by dividing \cref{eq:eta_gamma} by \cref{eq:gamma_HR}. Under this definition of $\Gamma$, $\eta \propto \eta_\subI$ and $\eta \propto \eta_\subO$. Losses within the interferometer affect $\Gamma(\Omega)$ slightly, and  $\eta \propto \eta_\subR$ is only approximate. \Cref{sec:radiation_pressure_calc} gives an example of how losses affect $\eta$ and $\Gamma$. The primary alternative definition is to use $\Gamma = N\big|_{S=1}$, but this definition makes $\eta_\subO$ both less physically intuitive and also sensitive to interferometer parameters.

Subtracting $\eta\Gamma$ from \cref{eq:N_no_sqz_derivation} and factorizing by the optical paths provides the definition of the remaining efficiency terms.
\begin{align}
  (1 - \eta_\subO)
  &= 
  \sum_{i}\left|\TLO \Tmat{T}_{\text{O},\smu}\right|^2
    \\
  \eta_\subO(1 - \eta_\subR) \Gamma
  &= 
  \sum_{i}\left|\TLO \Tmat{T}_{\text{R},\smu}\right|^2
    \\
  \eta_\subO\eta_\subR(1 - \eta_\subI) \Gamma
  &= 
  \sum_{i}\left|\TLO \Tmat{T}_{\text{I},\smu}\right|^2
\end{align}
Which add together to create the loss term in \cref{eq:metric_N}.
\begin{align}
    \Lambda_\IRO\Gamma &= \eta_\subO\eta_\subR(1 - \eta_\subI)\Gamma
    + \eta_\subO(1 - \eta_\subR)\Gamma
    + (1 - \eta_\subO)
\end{align}

\subsection{Ideal Interferometer Example}
\label{sec:ideal_IFO_example}

The derivations are now extended to recreate and generalize the ideal noise model of \cref{sec:ideal_IFO_metrics}, using \cref{eq:Kchi_standard} for $\K$. The two-photon matrix corresponding to the interferometer in \cref{fig:SQZ_mm_cavities} is given below for the lossless interferometer that is perfectly on resonance.
\begin{align}
  \Tmat{H}_\subR(\Omega) &\simeq
  \begin{bmatrix}
    \tfr(\Omega) & 0\\
    \K(\Omega) & \tfr(\Omega)
  \end{bmatrix},
  &
  \tfr(\Omega) &\simeq
\frac{\gamma_\Arm - i\Omega}{\gamma_\Arm + i\Omega},
                 &\Tmat{\Lambda}_\subR &= \Tmat{0}
\end{align}
In the ideal lossless case, the input and output paths also have perfect efficiency $\eta_\subI \simeq 1$ with $\Tmat{H}_\subI(\Omega) = \eta_\subI\Tmat{1},\; \Tmat{\Lambda}_\subI(\Omega) = \sqrt{1-\eta_\subI}\Tmat{1}$ and similarly for the output. These can be used to compute $\Tmat{H}$ and $\vec{m}$.
\begin{align}
  m_q &= \cos(\zeta)\K(\Omega) + \sin(\zeta)\tfr(\Omega)
  ,&
  m_p &= \cos(\zeta)\tfr(\Omega)
\end{align}
The equations above maintain the correct phase information for this ideal case analysis. Interestingly, $\K$ and $\tfr(\Omega)$ have different magnitude responses resulting from different factors of $\gamma_\Arm \pm i\Omega$, yet their phase response is the same. This Kramers-Kronig coincedence ensures $\Xi(\Omega)=0$ as long as the $\chi(\Omega)$ contribution to $\K(\Omega)$ is purely real. Thus, lossy mechanics will cause QRPN to dephase injected squeezing. This will not happen to any meaningful level for LIGO, but is noteworthy for optomechanics experiments operating on mechanical resonance.

The $\vec{m}$ above also includes the effect of the readout angle. For $\zeta=0$, it recovers \crefrange{eq:optical_gain_g}{eq:Kchi_standard}. More generally, it gives
\begin{align}
  \Gamma(\Omega) &= |\cos(\zeta)\K(\Omega)|^2 + \sin(2\zeta)|\K(\Omega)| + 1
  \\
  \theta(\Omega) &= \arctan\big(|\K(\Omega)| - \tan(\zeta)\big)
\end{align}
The exact expressions above can be simplified to better relate them to the LIGO data. Firstly, the squeezing angle is modified to be 0 at high frequencies, to match the conventions of the data. This modified angle is $\theta'(\Omega) = \theta(\Omega) - \theta(\Omega{\gg}\gamma_\Arm)$. Secondly, small shifts of the homodyne angle are linearized.
\begin{align}
  \Gamma'(\Omega) &\approx
                   \big(1 - |K(\Omega)|\big)^2 + 2\big(1 + \zeta\big)|K(\Omega)|
  \\
  \theta'(\Omega) &\approx \arctan\big(|\K(\Omega)|) - \zeta \frac{|\K(\Omega)|^2}{1 + |\K(\Omega)|^2}
\end{align}
The linearized $\Gamma'$ shows that , when $\zeta=-13^\circ=-0.23$, for frequencies near $\Omega_\sql$, $\K(\Omega_\sql) \simeq 1$, the interferometer quantum noise is reduced by about 23\% with respect to a nominal $\xi=0$ readout. This change is shown in the blue vs. grey plotted data for the Livingston loss plot in \cref{fig:data_Q} of $1-\eta$. There, $\eta$ changes as the $\Gamma$ model changes since only $\eta\Gamma$ can be measured due to subtracting an unsqueezed reference dataset. The $23\%$ noise reduction corresponds to approximately 1dB improvement from pondermotive quantum correlations. The angle formula above indicates that for frequencies $\Omega\lesssim\Omega_\sql$, the local oscillator also adds some additional shift to $\theta$ at low frequency, which is also observed in the LLO angle fits.

This analysis gives an example of how the derivations of this section are applied to extend the existing ideal interferometer models towards the real instruments. Exact models including more optical physics are yet more analytically opaque, but give a more complete complete picture if implemented numerically. \Cref{sec:matrix_coupled_cavity} shows the full matrix solution, including the cavities, to recover these equations while also handling cavity length offset detunings. It also includes transverse modal mismatch in its description. \Cref{sec:radiation_pressure_calc} gives the minimal extension of this ideal lossless interferometer to incorporate transverse mismatch, showing how the noise gain, $\Gamma$,  and rotation angle $\theta$ change specifically from mismatch. In particular, it shows that relating a measurement of $\Omega_\sql$ using squeezing back to the arm power $P_\Arm$ using \cref{eq:optical_gain_g} and  \cref{eq:Kchi_standard} is biased by transverse mismatch.

\section{Cavity Modeling and Metrics}
\label{sec:modeling}

The previous section derives the general form of the squeezing metrics using matrices of the two photon formalism.
For passive systems, the optical transfer function, $\tfh(\Omega)$, given at every sideband frequency, is sufficient to characterize the response to externally-supplied squeezing. The conceptual simplification and restriction to using only transfer functions is useful for interferometer modeling. Transfer functions, being complex scalar functions, are suitable for analytic calculations of cavity response and can be decomposed into rational function forms to inspect the rational roots, zeros and poles, and the overall gain of the response.

This section analyzes the coupled cavity system of the interferometer, depicted in \cref{fig:SQZ_mm_cavities}, through its decomposition into roots. More complicated transverse modal simulations analyze the frequency response of the interferometer cavities for each optical mode to every other mode. Modal simulations thus output a matrix of transfer functions, $\mat{H}(\Omega)$, which is difficult to analytically manipulate, but \cref{sec:modeling_TMM} shows how it can be projected back to a single scalar transfer function $\tfh(\Omega)$ and further simplified into the squeezing metrics.

The transfer function techniques of this section elucidate new squeezing results by avoiding the combined complexity of both two-photon and modal vector spaces. The full generality of two-photon matrices is only required for active systems that introduce internal squeezing, parametric gain or radiation pressure. Passive systems have the property that $\hat{q}_\text{out}$, $\hat{p}_\text{out}$ also obey the expectations of \cref{eq:vacuum_expectations,eq:vacuum_expectations2}. Following the notation of \cref{sec:derivations}, this results in the following condition.
\begin{align}
  \Tmat{1} &= \Tmat{H}\Tmat{H}^\dagger
  +\sum_{\smu}\Tmat{T}_{\smu}\Tmat{T}^\dagger_{\smu}
             \label{eq:passivity_condition}
\end{align}
Additionally, $\Gamma = 1$ is implied by that condition.
Without parametric gain, photons at upper and lower sideband frequencies are never correlated by a passive system. By the passivity condition and manipulations between sideband and quadrature basis, \cref{sec:passive_derivations} derives the squeezing metrics purely in terms of the transfer function $\tfh(\Omega)$.
\begin{align}
  \theta(\Omega)
  &=
    \big(\arg\big(\tfh(+\Omega)\big) + \arg\big(\tfh(-\Omega)\big)\big)/2
    \label{eq:theta_passive}
  \\
  \eta(\Omega)
  &=
    \big( |\tfh(+\Omega)|^2 + |\tfh(-\Omega)|^2\big)/2
    \label{eq:eta_passive}
  \\
  \Xi(\Omega)
  &=
    \big( |\tfh(+\Omega)| - |\tfh(-\Omega)| \big)^2/4\eta
    \label{eq:Xi_passive}
\end{align}
Quantum filter cavities are a method to use an entirely optical system to reduce the radiation pressure associated with squeezed light\cite{McCuller-PRL20-FrequencyDependentSqueezing, Zhao-PRL20-FrequencyDependentSqueezed}. They are passive cavities, and provide a useful example to study these squeezing metric formulas.
The first of these, \cref{eq:theta_passive}, is a well-established formula for the filter cavity design. It indicates that for cavities with an asymmetric phase response, usually due to being off-resonance or ``detuned'', that the squeezing field picks up a frequency-dependent quadrature rotation. Such a rotation applied in $\Tmat{H}_\subI$ can be generated by a cavity with transfer function $\tfh_\subI(\Omega)$ before the interferometer. This cavity rotation compensates the $\theta(\Omega)$ due to $\Tmat{H}_\subR$. Together, the product $\Tmat{H}_\subR\Tmat{H}_\subI$ has $\theta(\Omega) = 0$, allowing a single choice of squeezing angle $\phi$ to optimize $N(\Omega)$ at all frequencies.

The formulas \cref{eq:eta_passive} and \cref{eq:Xi_passive} indicate how losses represented in a transfer function translate to loss-like and dephasing degradations from cavity reflections. For filter cavities, these degradations are investigated in \cite{Kwee-PRD14-DecoherenceDegradation}, but this new factorization into scalar functions clarifies the discussion. The efficiency $\eta(\Omega)$ behaves as expected, an average of the loss in each sideband. The form of $\Xi(\Omega)$ is less expected, showing how the combination of loss and detuning in filter cavities creates noise that scales with the squeezing level. A simple picture for the dephasing effect is that when optical quadratures are squeezed, the noise power in both upper and lower sidebands is strictly increased. The sideband correlations allow the increased noise to subtract away for squeezed quadrature measurements but to add for measurements in the anti-squeezed quadrature. The asymmetric losses of detuned cavities preserve the noise increase on one sideband, while degrading the correlations. This ruins the subtraction for the squeezed quadrature and introduces $\Xi(\Omega) > 0$. This source of noise is $e^{\pm r}$ squeezing level dependent but entirely unrelated to fluctuations of the squeezing phase $\phiRMS$.

\subsection{Single Cavity Model for Interferometers}
\label{sec:modeling_cavities}

This section analyzes the effect of the interferometer cavities on squeezing. It starts by considering an interferometer with only one cavity - either in the Michelson arms or from a mirror at the output port, but not both. It represents the first generation of GW detectors. This single cavity scenario is also similar to a quantum filter cavity, in the regime of small detuning\cite{Komori-PRD20-DemonstrationAmplitude, Corbitt-PRD04-OpticalCavities, Khalili-PRD08-IncreasingFuture}. Advanced LIGO uses a coupled cavity system, depicted in \cref{fig:SQZ_mm_cavities}, and the transfer function equations for the reflection from the resonant sideband extraction cavity is extended in the next subsection to include the loss and detuning of the additional cavity.

A single cavity operated near resonance may be described using the scale parameters of the cavity bandwidth $\gamma_\Arm$, loss rate $\lambda_\Arm$ and detuning frequency $\delta_\Arm$, which are computed from the physical parameters of the mirror transmissivity $T_\itm$, round-trip loss $\Lambda_\arm$, cavity length $L_\arm$, and microscopic length detuning $\Delta L_\arm$.
\begin{align}
  \gamma_\Arm  &= \frac{c T_\itm}{4 L_\arm}
  & 
  \lambda_\Arm &= \frac{c \Lambda_\arm}{4 L_\arm}
  &
  \delta_\Arm  &= -ck\frac{\Delta L_\arm}{L_\arm}
                 \label{eq:cavity_relations}
\end{align}
These relations are accurate in the high-finesse limit $T_\itm \ll 1$, and combine to give the transfer function of the frequency-dependent cavity reflection.
\begin{align}
  \tfr_1(\Omega)
  &\approx
    -\frac{(\gamma_\Arm - \lambda_\Arm) - i(\Omega - \delta_\Arm)}{(\gamma_\Arm + \lambda_\Arm) + i(\Omega - \delta_\Arm)}
    \label{eq:single_cavity}
\end{align}
Notably, the sign of the reflectivity for a high-finesse cavity on resonance $\tfr_1(\Omega \ll \gamma_\Arm)=-1$,
but outside of resonance $\tfr_1(\Omega \gg \gamma_\Arm)=1$. This sign determines constructive or destructive interference in transverse mismatch loss analyzed in the next section. The internal losses of the cavity $\Lambda_\arm$ become cavity-enhanced in the reflection, causing squeezing to experience losses of $\Lambda_\Arm$.
\begin{align}
  \Lambda_\Arm
  &\equiv 1 - \eta(\Omega)\bigg|_{\substack{\tfh = \tfr_1\\ |\Omega| \ll \gamma_\Arm}}
  \approx \frac{4\lambda_\Arm}{\gamma_\Arm}
                            \approx \frac{4\Lambda_\arm}{T_\itm}
\end{align}
Furthermore, detuning the cavity off of resonance causes a rotation of reflected squeezing. For small detunings, the rotation can be approximated.
\begin{align}
  \theta(\Omega)\bigg|_{\substack{\tfh = \tfr_1\\ k\Delta L_\arm \ll T_\itm}}
  &\approx
  \frac{2\delta_\Arm\gamma_\Arm}{\gamma_\Arm^2 + \Omega^2}
    \approx
   -k \Delta L_\arm\frac{8}{T_\itm}\frac{\gamma_\Arm^2}{\gamma_\Arm^2 + \Omega^2}
\end{align}
Fluctuations in $\Delta L_\arm$ or $\delta_\Arm$ lead to a phase noise analogous to $\phiRMS$, but with the frequency dependence from the above equation\cite{Kwee-PRD14-DecoherenceDegradation}. Additionally, losses in the cavity lead to intrinsic dephasing $\Xi(\Omega)$, calculated below. This calculation is valid at any detuning $\delta_\Arm$, even those larger than the cavity width $\gamma_\Arm$. Its validity only requires being in the overcoupled cavity regime, where losses $\lambda_\Arm \lesssim \gamma_\Arm/2$.
\begin{align}
  \Xi(\Omega)\bigg|_{\substack{\tfh = \tfr_1}}
  &\approx
    \left(
  \frac{4\gamma_\Arm\lambda_\Arm\delta_\Arm\Omega}{\left( \gamma_\Arm^2 + (\Omega - \delta_\Arm)^2 \right)\left( \gamma_\Arm^2 +  (\Omega + \delta_\Arm)^2 \right)}\right)^2
\end{align}
When plotted, this expression for $\Xi(\Omega)$ has a Lorentzian-like profile, with a peak at $\Omega_{\Xi\text{max}}$. Above $|\delta_\Arm| \gtrsim \gamma_\Arm$, where the cavity resonance acts entirely either on upper or lower sidebands, the peak dephasing reaches a maximum. At small detunings, $|\delta_\Arm| \lesssim \gamma_\Arm$, the sideband loss asymmetry scales with the detuning.
\begin{align}
  \Omega_{\Xi{\text{max}}} &\approx \sqrt{\gamma^2_\Arm/4 + \delta^2 },
  & \Xi_{\text{max}} 
  &\approx 
    \frac{\lambda^2_\Arm}{\gamma^2_\Arm}{\cdot}\frac{8\delta^2_\Arm}{5\gamma^2_\Arm + 8\delta^2_\Arm}
    \label{eq:detuning_dephasing}
\end{align}
This single cavity model is also useful for analyzing quantum filter cavities and, like the $\Xi$ metric itself, these peak values have not been calculated in past frequency-dependent squeezing work. Conventional squeezing phase uncertainty, $\phiRMS$, can be cast into the units RMS radians of phase deviation, leading to the noise suppression limit for squeezing $S(\Omega)\ge 2\phiRMS$, by \cref{eq:N_limit}. For highly detuned cavities such as quantum filter cavities, $\sqrt{\Xi(\Omega)} \approx \Lambda_\subfc / T_\subfc$. Using the parameters of the A+ filter cavity \cite{Whittle-PRD20-OptimalDetuning}, $\Lambda_{\subfc}{\approx}60\text{ppm}$ and $T_{\subfc} {=} 1000 \text{ppm}$ indicates that optical dephasing is of order $60\text{mRad}$. For an optimal filter cavity with low losses\cite{Whittle-PRD20-OptimalDetuning}, this dephasing maximum occurs at $\Omega_{\Xi{\text{max}}} = \sqrt{5/8}\Omega_{\text{SQL}}$. This level of dephasing is commensurate with or even exceeds the expected residual phase uncertainty $\phiRMS < 30\text{mRad}$.

Optical dephasing from the LIGO interferometer cavities is not expected to be large for as they are stably operated on-resonance; however, detuned configurations of LIGO\cite{Ganapathy-PRD21-TuningAdvanced} are limited by dephasing from the unbalanced response and optical losses in the signal recycling cavity.

\subsection{Double Cavity Model for Interferometers}

For interferometers using resonant sideband extraction, like LIGO, the arm cavities have a length $L_\arm$, an input transmissivity of $T_\itm$, and are each on resonance to store circulating laser power. The signal recycling cavity (SRC) has a length $L_\src$ and a signal recycling mirror (SRM) of transmissivity $T_\srm$. The SRM forms a cavity with respect to the arm input mirror that resonantly increases the effective transmissivity experienced by the arm cavities to be larger than $T_\itm$, broadening the signal bandwidth. While the SRC is resonant with respect to the arm input mirror, it is anti-resonant with respect to the arm cavity, due to the negative sign of \cref{eq:single_cavity}. The anti-resonance leads to the opposite sign in the reflection transfer function below, \cref{eq:double_cavity}. The discrepancy in resonance vs. anti-resonance viewpoints is why the signal recycling cavity is also called the signal extraction cavity in GW literature.

The coupled cavity forms two bandwidth scales for the system, $\gamma_\Arm$, the modified effective arm bandwidth, and $\gamma_\Src$, the bandwidth of the signal recycling cavity. The arm and signal cavities have their respective round-trip losses $\Lambda_\arm$ and $\Lambda_\src$, as well as length detunings $\Delta L_\arm$, $\Delta L_\src$. In practice, the arm length detuning is expected to be negligible to maximize the power storage, but the signal recycling cavity detuning can be varied by modifying a bias in the control system that stabilizes $\Delta L_\src$.

The scale parameters for the cavity transfer function are approximated from the physical parameters:
\begin{align}
  u_\itm &= 1 - \sqrt{1 - T_\itm}
  &
  u_\srm &= 1 - \sqrt{1 - T_\srm}
           \label{eq:u_factors}
  \\
  \gamma_\Arm  &= \frac{c u_\itm}{2 L_\arm} \cdot\frac{2 - u_\srm}{u_\srm}
  & 
    \gamma_\Src   &= \frac{c u_\srm}{2 L_\src}
                    \label{eq:double_cavity_gamma}
  \\
  \lambda_\Arm &= \frac{c}{L_\arm} \left(\frac{\Lambda_\arm}{4} - \frac{u_\itm \Lambda_\src}{2 u_\srm^2} + \frac{u_\itm \Lambda_\src}{4 u_\srm}\right)
  &
    \lambda_\Src  &= \frac{c\Lambda_\src}{4L_\src}\left(1 - \frac{u_\srm}{2} \right)
  \\
  \delta_\Arm &= -ck\frac{\Delta L_\arm}{L_\arm}  - \frac{\gamma_\Arm}{\gamma_\Src}\delta_\Src
  &
  \delta_\Src &= -ck\frac{\Delta L_\src}{L_\src} 
           \label{eq:delta_factors}
\end{align}
These approximations are valid for the LIGO mirror parameters, see \cref{tab:LIGO_params}, and model the loss and detuning to {\color{red} $5\%$} accuracy. They are derived in \cref{sec:scalar_coupled_cavity} from Taylor expansions, solving roots, and selectively removing terms. Expanding in the $u$ factors of \cref{eq:u_factors} gives lower error than expanding in transmissivity or reflectivity factors directly, due to the low effective finesse of the coupled cavity system and the high transmissivity of the SRM.
\begin{table}
\centering
\caption{Parameters of LIGO for data fitting and modeling}
\begin{ruledtabular}
\begin{tabular}{l|c|cc}
Parameter & Symbol & LLO Value & LHO Value \\
\hline
arm input transmissivity & $T_\itm$&0.0148 & 0.0142 \\
arm length & $L_\arm$ & \multicolumn{2}{c}{3995 m}\\
arm round-trip loss & $\Lambda_{\arm}$ & \multicolumn{2}{c}{${\sim}80$ppm}\\
SRM transmission& $T_\srm$&\multicolumn{2}{c}{0.325}  \\
SRC length & $L_\src$ &\multicolumn{2}{c}{55 m}\\
SRC round-trip loss & $\Lambda_\src$ & \multicolumn{2}{c}{$\lesssim 3000$ppm}\\
  \hline
Mirror mass& $m$ &  \multicolumn{2}{c}{39.9kg}\\
Arm power& $P_\Arm$ & $200{\pm} 10\text{ kW}$  &  $190{\pm}10\text{ kW}$ \\
QRPN crossover& $\Omega_{\sql}$ & $2\pi\cdot 33$ Hz & $2\pi\cdot 30$ Hz \\
  \hline
arm signal band& $\gamma_\Arm$& $2\pi\cdot450$ Hz &  $2\pi\cdot410$ Hz \\
SRC band& $\gamma_\Src$& \multicolumn{2}{c}{$2\pi\cdot80$kHz} \\
Arm length detuning & $\Delta L_{\arm}$ & \multicolumn{2}{c}{0nm}\\
SRC length detuning & $\Delta L_{\src}$ & 1.02nm & 1.23nm\\
  \hline
arm resonant loss& $\Lambda_\Arm$ & \multicolumn{2}{c}{$\lesssim 2000$PPM}\\
SRC resonant loss& $\Lambda_\Src$ & \multicolumn{2}{c}{${\sim} 1\%$ to $3\%$}\\
arm/SRC detuning& $\delta_{\Arm}$ & $2\pi\cdot 10.1$Hz & $2\pi\cdot 11.2$Hz \\
  \hline
Injected squeezing & $e^{\pm 2r}$ & {${\pm}9.7\text{ dB}$}  &  {${\pm}8.7\text{ dB}$} \\
SQZ-OMC mismatch& $\Upsilon_{\subO}$ & $2\%$ & $4\%$ \\
Reflection mismatch (fit)& $\Upsilon_{\subR}$ & $12\%$ & $35\%$ \\
Additional SQZ loss (fit)& $\Lambda_{\text{IO}}=1{-}\eta_\subI\eta_\subO$ & $31\%$ & $34\%$ \\
\end{tabular}
\end{ruledtabular}
\label{tab:LIGO_params}
\end{table}
The scale factors result in the following reflectivity transfer function.
\begin{align}
  \tfr_2(\Omega)
  &=
    \frac{(\gamma_\Arm - \lambda_\Arm) - i(\Omega - \delta_\Arm)}{(\gamma_\Arm + \lambda_\Arm) + i(\Omega - \delta_\Arm)}
    {\cdot}\frac{(\gamma_\Src - \lambda_\Src) - i(\Omega - \delta_\Src)}{(\gamma_\Src + \lambda_\Src) + i(\Omega - \delta_\Src)}
    \label{eq:double_cavity}
\end{align}
Notably, this reflectivity is $\tfr_2(\pm\Omega \ll \gamma)=1$ and $\tfr_2(\gamma_\Arm \ll \pm\Omega \ll \gamma_\Src)=-1$ which has an opposite overall sign to that of single cavity interferometers. On reflection, the squeezing field experiences different cavity enhanced losses depending on the frequency.
\begin{align}
  \Lambda_{\Src} &\equiv 
1 - \eta(\Omega)\bigg|_{\substack{\tfh = \tfr_2\\ \gamma_\Arm \ll |\Omega| \ll \gamma_\Src}}
&&\hspace{-3em}\approx \frac{2 - u_\srm}{u_\srm}\Lambda_\src
  \\
  \Lambda_{\Arm} &\equiv 
1 - \eta(\Omega)\bigg|_{\substack{\tfh = \tfr_2\\ |\Omega| \ll \gamma_\Arm}}
&&\hspace{-3em}\approx \frac{4\lambda_\Arm}{\gamma_\Arm}+\Lambda_\src \approx \frac{u_\srm}{u_\itm}\Lambda_\arm
\end{align}
The dataset of \cref{sec:experiment} shows frequency dependent losses, where the loss increases $12\%$ for LLO and $33\%$ for LHO. Assuming the losses result from the equations above, this corresponds to round-trip losses in the LIGO signal recycling cavities, $\Lambda_\src$, of $1.1\%$ to $3.2\%$, which is not realistic. Most mechanisms that introduce loss in the SRC would also introduce it into the power recycling cavity in an obvious manner. The current power recycling factors exclude this possibility, and independent measurements of $\gamma_\Arm$ bound $\Lambda_\src$ losses to ${\le} 3000$ppm. The next section investigates how transverse mismatch can result in this level of observed losses.

In addition to the losses, \cref{eq:double_cavity} can be used to determine the cavity-induced squeeze state rotation from the detuning of the signal recycling cavity.
\begin{align}
  \theta(\Omega)\bigg|_{\substack{\tfh = \tfr_2\\ \delta_\Src \ll \gamma_\Src \\ \Delta L_\arm = 0}}
  &\approx
  \frac{2\delta_\Arm\gamma_\Arm}{\gamma_\Arm^2 + \Omega^2}
    +
  \frac{2\delta_\Src\gamma_\Src}{\gamma_\Src^2 + \Omega^2}
    \\
  &\approx
    k\Delta L_\src\frac{4}{u_\srm}
    \left(
    \frac{\gamma^2_\Src}{\gamma_\Src^2 + \Omega^2}
    -
    \frac{\gamma^2_\Arm}{\gamma_\Arm^2 + \Omega^2}
 \right)
    \label{eq:double_cav_theta}
\end{align}
This indicates the surpising result that detuning the SRC length does not affect the squeezing within the effective
arm bandwidth to first order. Instead, it adds the squeezing rotation in the middle band above the arm bandwidth but below the SRC bandwidth. In the data analysis of \cref{sec:experiment} and \cref{fig:data_Q}, the convention for $\theta(\Omega)$ is set to be 0 at ``high'' frequencies in this intermediate cavity band, in which case it appears to cause a rotation around $\gamma_\Arm$. This convention used for the data corresponds to omitting the first, $\gamma_\Src$-scaled term of \cref{eq:double_cav_theta}.

\section{Transverse Mismatch Model}
\label{sec:modeling_TMM}

Squeezing, as it is typically implemented for GW interferometers, modifies the quantum states in a single optical mode. For LIGO, this mode is the fundamental Gaussian beam resonating in the parametric amplifier cavity serving as the squeezed state source. The cavity geometry establishes a specific complex Gaussian beam parameter that defines a modal basis decomposition into Hermite Gaussian (HG) or Laguerre Guassian (LG) modes. That basis is transformed and redefined during the beam propagation through free space and through telescope lenses on its way to and from the interferometer. The cavities of the interferometer each define their own resonating beam parameters and respective HG or LG basis of optical modes.

In practice, the telescopes propagating the squeezed beam to and from the interferometer imperfectly match the complex beam parameters, so basis transformations must occur that mix the optical modes. The mismatch of complex beam parameters is called here ``transverse mismatch''. Non-fundamental HG or LG transverse modes do not enter the OPA cavity, and so carry standard vacuum rather than squeezing. Basis mixing from transverse mismatch thus leads to losses; however, unlike typical losses such basis transformations are coherent and unitary, which leads to the constructive and destructive interference effects studied in this section.

The interferometer transfer function $\tfh(\Omega)$ is a single scalar function representing the frequency dependence of the squeezing channel from source to readout, but the optical fields physically have many more channels. The cavities visited by the squeezed states each have a transfer function matrix in their local basis, given by $\mat{H}_{\subI}$, $\mat{H}_{\subR}$, $\mat{H}_{\subO}$ for the squeezing input, interferometer reflection, and system output respectively. The diagonals of these matrices indicate the frequency response during traversal for every transverse optical mode. The off-diagonals represent the coupling response between modes that result from scattering and optical wavefront errors.

Between the cavities, $\mat{U}$ matrices represent the basis transformations due to transverse mismatch. Here, $\vec{e}_{\text{sqz}}$, $\vec{e}_{\text{read}}$ are basis vectors for projecting from the single optical mode of the emitted squeezed states and to the single mode of the optical homodyne readout defined by its local oscillator field.
\begin{align}
\tfh(\Omega) &= {\vec{e}_{\text{read}}}^{\,\dagger} \mat{H}(\Omega)\vec{e}_{\text{sqz}}
                  \label{eq:TF_projection}
  \\
\mat{H}(\Omega) &=\mat{H}_\subO\mat{U}_{\subO, \subR}\mat{H}_\subR\mat{U}_{\subR, \subI}\mat{H}_\subI
                  \label{eq:TF_matrix}
\end{align}
\Cref{eq:TF_projection} and \cref{eq:TF_matrix} give the general, basis independent, form to compose the effective transfer function for the squeezed field using a multi-modal simulation of a passive interferometer. This is complicated in the general case, but the following analysis develops a simpler, though general, model for how transverse mismatch manifests as squeezing losses.

Transverse mismatch is often physically measured as a loss of coupling efficiency, $\MM$, of an external Gaussian beam to a cavity measured as a change in optical power. Realistically, more than two transverse modes are necessary to maintain realistic and unitary basis transformations, but, for small mismatches of complex beam parameters, $\Upsilon < 10\%$. In this case, only the two lowest modes in the Laguerre-Gauss basis have significant cross-coupling. For low losses, the fundamental Gaussian mode, LG0, loses most of its power to the radially symmetric LG1 mode, assuming low astigmatism and omitting azimuthal indices. This motivates the following simplistic two-mode model to analyze the effect of losses on $\tfh(\Omega)$. In this model $\mat{U}$ gives the unitary, though not perfectly physical, basis transformation:
\begin{align}
  \mat{U}(\MM, \psi, \phi)
  &\equiv
    e^{i\phi}\begin{bmatrix}
    \sqrt{\cMM} & -e^{i\psi}\sqrt{\MM}
    \\
    e^{-i\psi}\sqrt{\MM} & \sqrt{\cMM}
  \end{bmatrix}
                           \label{eq:U_def}
\end{align}
This unitary transformation includes two unknown phase parameters. The first, $\psi$, is the phase of the mismatch, which characterizes whether beam size error or wavefront phasing error dominates the overlap integral of the external LG0 and cavity LG1 modes. The second, $\phi$, is the mismatch phase error from the external LG0 to the cavity LG0. The $\phi$ term is included above to fully express the unitary freedom of $\mat{U}$, but is indistinguishable from path length offsets, physically controlled to be $0$, and ignored in further expressions.
\autofiguresvgTEX{
  folder=./figures/, 
  file=SQZ_mm_chain, 
  caption={
    Propagation of the squeezed beam and unsqueezed higher order transverse beam modes from source to readout. The stages (a)-(d) correspond to the components in \cref{fig:SQZ_mm_IFO}, depicting the matrix math of \crefrange{eq:H_refl}{eq:h_chain}. $\MM_\subI$ represents the transverse mismatch loss of the squeezing to interferometer, and $\MM_\subO$ is the mismatch of the squeezing to readout via the output mode cleaner. These mismatches cause beamsplitter-like mixing between the LG0 and LG1+ modes through \cref{eq:U_def}. $\psi_\subI$,  $\psi_\subO$,  $\psi_{\text{G}}$ are unmeasured phasing terms of the interferometer and output mismatch and of the Gouy-phase advance from the beam propagating to the output mode cleaner.
  },
  label=SQZ_mm_chain,
}

In the case of a GW interferometer with an output mode cleaner, there are two mode matching efficiencies expressed as individually measurable parameters. The first is the coupling efficiency (in power) and phasing associated between the squeezer and interferometer $\MM_\subI, \psiI$. The second are parameters for efficiency and phasing between the squeezer and output mode cleaner, $\MM_\subO, \psiO$, which defines the mode of the interferometer's homodyne readout. Both cases represent a basis change from the Laguerre-Gauss modes of the squeezer OPA cavity into the basis of each respective cavity. In constructing $\mat{H}(\Omega)$, however, the squeezing is transformed to the interferometer basis, reflects, and then transforms back to the squeezing basis. This corresponds to the operations of \cref{fig:SQZ_mm_chain}. There are also parameters to express the coupling efficiency and phase, $\MM_\subF, \psiF$, between the interferometer cavity and the OMC cavity. The $\MM_\subF$ parameter is less natural to analyze squeezing is not independent from  $\MM_\subI$ and $\MM_\subO$. It is considered at the end of this section, as it can also be independently measured.

\cref{fig:SQZ_mm_chain} is implemented into \cref{eq:TF_matrix} through this simplistic two-mode representation by assuming that the interferometer reflection transfer function $\tfr(\Omega)$ applies to the LG0 mode in the interferometer basis. The LG1 mode picks up the reflection transfer function $\tfr_{\text{hom}}$, which is approximately ${\sim}1$ due to high order modes being non-resonant in the interferometer cavities and thus directly reflecting. %
\begin{align}
  \mat{H}_\subR &= 
  \begin{bmatrix}
    \tfr(\Omega) & 0
    \\
    0 & \tfrH(\Omega)
  \end{bmatrix},
  &
  \mat{G} &= 
    \begin{bmatrix}
    1 & 0
    \\
    0 & e^{i\psi_G}
  \end{bmatrix}
        \label{eq:H_refl}
        \\
  \tfr(\Omega) &= \tfr_2(\Omega) \text{ or } \tfr_1(\Omega)
  &
  \tfrH(\Omega) &= e^{i\theta_{\text{hom}}} \approx 1
\end{align}
The reflection term $\tfr(\Omega)$ can use either the single, \cref{eq:single_cavity}, or double,  \cref{eq:double_cavity}, cavity forms. LIGO, using resonant sideband extraction, uses $\tfr_2(\Omega)$. Frequencies where the reflection takes a negative sign will be shown to experience destructive interference from modal basis changes, increasing squeezing losses. The $\mat{G}$ matrix includes a phasing factor due to additional Gouy phase of higher-order-modes. This factor is degenerate with the mismatch phasings $\psi_\subI$ and $\psi_\subO$ in observable effects. These matrices are composed per \cref{fig:SQZ_mm_chain} to formulate the overall transfer function of the squeezed field.
\begin{align}
  \mat{H}(\Omega) &= 
\underbrace{\mat{U}(\MMO, \psiO)
\mat{G}
\mat{U}^\dagger(\MMI, \psiI)}_{\mat{U}_{\text{O,R}}}
\mat{H}_\subR
                    \underbrace{
\mat{U}(\MMI, \psiI)
                    }_{\mat{U}_{\text{R,I}}}
                    \label{eq:H_chain}
                    \\
  \tfh(\Omega) &= 
  \begin{bmatrix}
    1 \\ 0
  \end{bmatrix}^T
  \mat{H}(\Omega)
  \begin{bmatrix}
    1 \\ 0
  \end{bmatrix}
                    \label{eq:h_chain}
  ,
  \text{   and using }
\mat{H}_\subO = \mat{H}_\subI = \mat{1}
\end{align}
Ignoring intra-cavity losses and detunings, the two reflection forms $\tfr_1$, $\tfr_2$ can be simplified to give their respective transfer functions $\tfh_1$, $\tfh_2$.

For quantum noise below $\Omega < \gamma_\Src$, the double cavity reflectivity $\tfr_2(\Omega)$ behaves like a single cavity, using the $\gamma_\Arm$ of \cref{eq:double_cavity_gamma} and with the opposite reflection sign as \cref{eq:single_cavity}. 
\begin{align}
  \tfr_2(\Omega) &\approx +\frac{\gamma_\Arm - i\Omega}{\gamma_\Arm + i\Omega}
  &\Rightarrow&&
  \tfh_2(\Omega) &= \sqrt{\textstyle \cMMO}\frac{\gamma_\Arm - i\aFactor\Omega}{\gamma_\Arm + i\Omega}
                   \label{eq:h2_TMM}
                  \\
  \tfr_1(\Omega) &= -\frac{\gamma_\Arm - i\Omega}{\gamma_\Arm + i\Omega}
  &\Rightarrow&&
  \tfh_1(\Omega) &= \sqrt{\textstyle \cMMO}\frac{i\Omega - \aFactor\gamma_\Arm}{i\Omega + \gamma_\Arm}
\end{align}
Using the factor
\begin{align}
  \aFactor &\equiv 1 - 2\MMI + 2\bFactor\sqrt{\MMI\MMO}e^{i\psiR}
             \intertext{where:}
  \bFactor &\equiv \textstyle\sqrt{\frac{\cMMI}{\cMMO}} \approx 1
             \\
  \psiR &\equiv \psiO + \psi_G - \psiI
\end{align}
The phasing factor $\psiR$ shows that the unknown mismatch phasings combine to a single unknown overall phase. This overall phase determines the extent to which the separate beam mismatches of $\MM_\subI$ and $\MM_\subO$ coherently stack or cancel with each-other. The factor $\aFactor$ is the total squeezer LG0 to readout LG0 coupling factor for the effective mode mismatch of the full system, specifically when the interferometer reflection $\tfr(\Omega) = -1$. As an effective mismatch, it can be related back to the diagonal elements of \cref{eq:U_def} to give an effective mismatch loss on reflection, $\MM_\subR$.
\begin{align}
  \MMR &= 1 - \textstyle |\aFactor|^2 \approx 4\MMI - 4\bFactor\sqrt{\MMI\MMO}\cos(\psiR)
         \label{eq:MMR_def}
\end{align}
This effective mismatch loss becomes apparent after computing the full system efficiency $\eta(\Omega)$ (\cref{eq:eta_passive}) using $\tfh_1$ and $\tfh_2$.
\begin{align}
  \eta_\subR(\Omega)\bigg|_{\tfh = \tfh_2}
  &=
    \left( \cMMO \right)\frac{\gamma_\Arm^2 + \left(\cMMR\right)\Omega^2}{\gamma_\Arm^2 + \Omega^2}
    \label{eq:effective_MMR_loss2}
                  \\
  \eta_\subR(\Omega)\bigg|_{\tfh = \tfh_1}
  &=
    \left( \cMMO \right)\frac{\Omega^2 + \left(\cMMR\right)\gamma_\Arm^2}{\gamma_\Arm^2 + \Omega^2}
    \label{eq:effective_MMR_loss1}
\end{align}
For the double cavity system of LIGO, \cref{fig:data_Q} is presented using the loss rather than efficiency. To relate to the measurement, the loss attributable to mode mismatch is then written
\begin{align}
  \Lambda_\MM(\Omega) \equiv 1 - \eta_\subR\bigg|_{\tfh = \tfh_2}
  &\approx
    \MMO + \frac{\Omega^2}{\gamma_\Arm^2 + \Omega^2}\MMR
    \label{eq:Lambda_MM}
\end{align}
Mode mismatches between the squeezer and OMC were directly measured during the LIGO squeezer installation to be $2\%-4\%$, and mismatches from the squeezer and interferometer were indirectly measured but are expected to be of a similar level. The large factors in \cref{eq:MMR_def} indicate that the independent mismatch measurements are compatible with the observed frequency dependence and levels of the losses to squeezing. The effective mismatch loss $\MMR$ has the following bounds with respect to the independent mismatch measurements.
\begin{align}
  \MMR &\approx 4\MMI &&\text{ when } \MMO = 0
                          \\
  0 \le \MMR &\le 8\MMI &&\text{ when } \MMI = \MMO
                          \\
  \MMR &\approx 4\MMI &&\text{ when averaged over } \psiR
                         \label{eq:TMM_bound_avg}
\end{align}
It is worth noting here how the realistic interferometer differs from this simple two-mode model. The primary key difference is that real mismatch occurs with more transverse modes. Expanding this matrix model to include more modes primarily adds more $\cos(\psiR)$-type factors to the last term of \cref{eq:MMR_def}. These factors will tend to average coherent additive mismatch between the squeezer and the OMC away, leaving only the squeezer to interferometer terms. Additionally, not only is there beam parameter mismatch from imperfect beam-matching telescopes, but there is also some amount of misalignment, statically or in RMS drift. Mismatch into modes of different order picks up different factors of $\psiG$. Together, including more modes leaves the bounds above intact, but makes \cref{eq:TMM_bound_avg} more representative given the expanded dimensionality of mismatch-space to average away $\cos(\psiR)$.

The other notable difference in realistic instruments is that the high order modes pick up small phase shifts of reflection, as the cavities are not perfectly out of resonance at all high order modes. This corresponds to $\tfr_{\text{hom}} \ne 1$. The signal recycling mirror is sufficiently low transmissivity that the finesse is low and, even when off-resonance, higher order modes pick up a small but slowly varying phase shift. This has the property of mixing the frequency dependent losses resulting from $\tfh_1$ and $\tfh_2$, resulting in a slightly more varied frequency-dependence that is captured in the full model of \cref{sec:matrix_coupled_cavity}.

While the phasing of the mismatch, $\psiR$, is not directly measurable, it manifests in an observable way. It adds to the complex phase of $\aFactor$ to cause a slight rotation of the squeezing phase, making the cavity appear as if it is detuned. the frequency dependence and magnitude of this rotation is given by (c.f. \cref{eq:theta_passive}),
\begin{align}
  \theta_\MM(\Omega) \equiv \theta\bigg|_{\tfh = \tfh_2}
  &\approx
    \frac{-\Omega^2}{\gamma_\Arm^2 + \Omega^2}2\bFactor\sqrt{\MMI\MMO}\sin(\psiR)
\end{align}
which adds to the rotation from cavity length detuning \cref{eq:double_cav_theta}. The addition of this term with $\sqrt{\MMI \MMO}$ unknown confounds the ability to use the data of \cref{fig:data_Q} to constrain $\psiR$. There is a small discrepancy between the length-detuning induced optical spring observed in the interferometer calibration \cite{Cahillane-PRD17-CalibrationUncertainty, Sun-CQG20-CharacterizationSystematic} and the detuning inferred from the data. The additional mismatch phase shift helps explain that such a discrepancy is possible, but the two should be studied in more detail. Note that the small Gouy phase shift from $\tfr_{\text{hom}}$ can be significant for this small detuning effect. The expression above is primarily provided to indicate the magnitude of variation as a function of $\sin(\psiR)$, so that future observations can better constrain $\psiR$ by comparing squeezing measurements of $\theta(\Omega)$ with calibration measurements of the optical spring arising from $\delta_\Src$.

The asymmetric contribution of $\aFactor$ in \cref{eq:h2_TMM} also causes mode mismatch to contribute to optical dephasing, $\Xi(\Omega)$ (c.f. \cref{eq:Xi_passive}). The dependence on $\tfr_{\text{hom}}, \MMI, \psiR$ is complex and does not have single dominating contributions, so an analytic expression is not computed here. Using the exact models of \cref{sec:matrix_coupled_cavity}, that mimic the datasets of \cref{sec:experiment} give a contribution of $\sqrt{\Xi}$ that peaks at $\gamma_\Arm$ and is 10-20\text{mRad} for the Livingston LLO model, and 10-50mRad for the Hanford LHO model, with a range due to imperfect knowledge of the mismatch parameters.

The transverse mismatch calculations so far use the parameters $\MMO$, which is directly measurable, and $\MMI$, which is independent, but $\MMI$ can not easily be measured using invasive direct measurements due to the fragile operating state of the GW interferometer. Another mismatch parameter exists for the signal beam traveling with the Michelson fringe-offset light. This beam experiences a separate mode matching efficiency, $\MMF$, denoting the mismatch loss between the interferometer and the OMC. $\MMF$ can be calculated from the original parameters by following the red signal path depicted in \cref{fig:SQZ_mm_chain}.
\begin{align}
e^{i\phi_\subF}
\mat{U}(\MMF, \psiS)
  &=
\mat{U}(\MMO, \psiO)
\mat{G}
\mat{U}^\dagger(\MMI, \psiI)
\end{align}
Expanding this form results in the following relations
\begin{align}
  \MMF &\approx \MMO + \MMI - 2\sqrt{\MMO\MMI}\cos(\psiR)
         \\
   \MMF &\approx \MMR/2 + \MMO - \MMI
\end{align}
Experimentally, $\MMF$ can be determined or estimated more directly than $\MMI$ by using signal fields from the arms, though can be confused with projection loss when the local oscillator readout angle $\xi\ne 0$ (c.f. \cref{eq:homodyne_observable}). These formulas provide the set of relations to estimate each of the mode mismatch parameters from the others, and potentially the overall mismatch phase $\psiR$ as well. These relations are calculated using the assumptions of this section: the two-mode approximation and that $\tfr_{\text{hom}} \simeq 1$.

Together, the relations of this section give insight in to how the physical mismatch parameters, $\MMI$, $\MMO$ and $\MMF$ contribute to squeezing degradations. $\MMR$ is a new form of effective mismatch parameter that is directly measurable from squeezing data, using the analysis of \cref{sec:experiment}. It indicates how squeezing changes with frequency due to \crefrange{eq:effective_MMR_loss2}{eq:effective_MMR_loss1}. Together, the complex, coherent interactions of transverse modal mixing on squeezed state can be concisely characterized in cavity-enhanced interferometers.

\subsection{Implications for Frequency Dependent Squeezing}
This analysis of the transverse mismatch applies to the reflection of squeezing off of any form of cavity. Namely, the detuned filter cavity for frequency-dependent rotation of squeezing in the LIGO A+ upgrade. This cavity will be installed on the input, $\Tmat{H}_\subI$ section of the squeezing transformation sequence. The filter cavity mismatch loss $\MM_{\subfc}$ will behave analogously to $\MMI$, introducing losses of ${\sim}4\MM_{\subfc}$ at frequencies resonating in the cavity. The mismatch loss adds to those caused by the internal round-trip cavity loss $\Lambda_{\subfc}$, creating the effective loss $\Lambda_{\subFC} \approx 4\MM_{\subfc} + 4\lambda_{\subFC}/\gamma_{\subFC}$ using \cref{eq:cavity_relations}.

The intra-cavity losses then set the scale for how much transverse mismatch is allowable before mismatch dominates the squeezing degradation, $\MM_{\subfc} < \Lambda_\subfc / T_\subfc$. More importantly, they add to the dephasing from the detuned cavity, by creating an effective $\lambda'_\subFC = \lambda_{\subFC} + \MM_{\subfc}\gamma_{\subFC}$ which can be used in \cref{eq:detuning_dephasing}. The dephasing will set the limit to the allowable injected squeezing $e^{\pm 2r}$ level as it introduces anti-squeezing at critical frequencies in the spectrum for astrophysics.

\section{Conclusions}

Before this work, the squeezing level in the LIGO interferometers was routinely estimated using primarily high-frequency measurements. This was done to utilize a frequency band where the classical noise contributions were small, while also giving a large bandwidth over which to improve the $\Delta F \Delta t$ statistical error in noise estimates. In doing so, LIGO recorded a biased view of the state of squeezing performance between the two instruments. The data analysis of this work has revealed several critical features to better understand and ultimately improve the quantum noise in LIGO.

First, it indicates that the two sites have similar optical losses in their injection and readout components, as seen from the low-frequency losses of \cref{fig:data_Q}. There is still a small excess of losses over the predictions given in \cite{Tse-PRL19-QuantumEnhancedAdvanced}, but substantially less than implied when estimating the losses using high frequency observations. The most culpable loss components in the LIGO interferometers are being upgraded for the next observing run.

Second, this data analysis indicates that squeezing is degraded particularly at high frequencies, and the modeling and derivations provide the mechanism of transverse optical mode mismatch, external to the cavities, as a plausible physical explanation. This will be addressed in LIGO through the addition of active wavefront control to better match the beam profiles between the squeezer's parametric amplifier, new filter cavity installation, interferometer, and output mode cleaner.

Third, the quantum radiation pressure noise is now not only measured, but employed as a diagnostic tool along with squeezing. QRPN indicates that the effective local oscillator angle in the Michelson fringe offset light at LLO is a specific, nonzero, value. This indicates that to power up the detector further, while maintaining a constant level of fringe light, the angle will grow larger and cause more pronounced degradation of the sensitivity by projecting out of the signal's quadrature. Ultimately, the LO angle should become configurable using balanced homodyne detection, another planned upgrade as part of ``A+''.

Finally, this work carefully derives useful formula to manipulate the quantum squeezing response metrics. These are useful to reason and rationalize the interactions of squeezing with ever more complex detectors, both for gravitational-wave interferometers, and more generally as squeezing-enhanced optical metrology becomes more commonplace. The design of a future generation of gravitational wave detectors must be optimized specifically to maintain exceptional levels of squeezing compared to today. The quantum response metrics derived in this paper will aid that design work by simplifying our interpretation of squeezing with simulations. With these diagnostics and the data from observing run 3, LIGO is now better prepared to install and characterize frequency dependent squeezing in its ``A+'' upgrade not as a demonstration, but for stable, long-term improvement of the quantum enhanced observatories to detect astrophysical events.

\begin{acknowledgments}
 LIGO was constructed by the California Institute of Technology and Massachusetts Institute of Technology with funding from the National Science Foundation, and operates under Cooperative Agreement No. PHY-0757058. Advanced LIGO was built under Grant No. PHY-0823459. The authors gratefully acknowledge the National Science Foundation Graduate Research Fellowship under Grant No. 1122374. 
\end{acknowledgments}

\bibliography{fdep_main}

\appendix
\section{Dephasing in Active Interferometers}
\label{sec:phase_noise_composition}

This appendix provides the technical derivation of \ref{eq:theta_calculation} and \ref{eq:Xi_calculation} in \cref{sec:derivations}, and uses the terms defined there. This derivation produces the intermediate steps in the relation \cref{eq:decomposition_relation}, starting from the right-hand-side of that equation. From there, the $\vec{m}$ effective observation vector can be inserted. This vector is complex, while the left-acting matrices $\Tmat{R}$ and $\Tmat{S}$ are both real. The final noise expression uses a vector norm that takes the square sum of all of the real and imaginary parts of the resulting vector. The vector norm can formally be replaced by a matrix Frobenius norm, notated $|\cdot|_{\text{F}}$, while the complex vector $\vec{m}$ is split into the real matrix $\Tmat{q}$.
\begin{align}
  \left|\TLO \Tmat{H}\Tmat{R}(\phi) \Tmat{S}(r)  \right|^2
  &=
    \left|
    \begin{bmatrix}
      \Mq \\ \Mp
    \end{bmatrix}^T
    \Tmat{R}(\phi) \Tmat{S}(r)  \right|^2
  =
    \left|
    \Tmat{q}^T \Tmat{R}(\phi) \Tmat{S}(r)  \right|^2_{\text{F}}
  \label{eq:noise_in_q}
\end{align}
\begin{align}
  \Tmat{q} &\equiv
    \begin{bmatrix}
      \Re \{\Mq\} &
      \Im \{\Mq\}
      \\
      \Re \{\Mp\} &
      \Im \{\Mp\}
    \end{bmatrix}
\end{align}
The $\Tmat{q}$ matrix can then undergo a singular value decomposition into two rotations acting on a real diagonal matrix. 
\begin{align}
\Tmat{R}(\theta_D)
    \begin{bmatrix}
      \Sigma_+ & 0 \\
      0 & \Sigma_- \\
    \end{bmatrix}\Tmat{R}(\theta_C)
                    &\equiv \Tmat{q}
\end{align}
The rotations are labeled $\theta_\text{D}$ and  $\theta_\text{C}$ for the differential and common rotations. The common angle expresses the average phase on both optical quadratures, physically due to transmission or cavity delay, whereas the differential angle expresses the rotation of the principle squeezing axis into a specific optical quadrature. $\theta_\text{D}$ calculated from the SVD is the exact form of \cref{eq:theta_calculation}. The decomposition may then be inserted into \cref{eq:noise_in_q} to create a scalar expression taking the form of the left-hand-side of \cref{eq:decomposition_relation}.
\begin{align}
  \left|
  \Tmat{q}^T \Tmat{R}(\phi) \Tmat{S}(r)  \right|^2_{\text{F}} 
  &=
    \left(\Sigma_+^2e^{-2r} + \Sigma_-^2e^{+2r}\right)
    \cos^2(\phi - \theta_D)
    \nonumber\\&\hspace{1em}
    +
    \left(\Sigma_+^2e^{+2r} + \Sigma_-^2e^{-2r}\right)
    \sin^2(\phi - \theta_D)
\end{align}
From there, terms can be extracted to form the relations of \crefrange{eq:metric_N}{eq:metric_Lambda}
\begin{align}
  \eta\Gamma &= \left| \TLO \Tmat{H} \right| = |\Mq|^2 + |m_b|^2 = \Sigma_+^2 + \Sigma_-^2
               \label{eq:eta_gamma_SVD}
               \\
\Xi \eta\Gamma &= \Sigma_-^2 \hspace{2.6em}
(1 - \Xi)\eta\Gamma = \Sigma_+^2 
               \label{eq:eta_gamma_SVD2}
\end{align}
Finally, dividing \cref{eq:eta_gamma_SVD2} by \cref{eq:eta_gamma_SVD} gives the dephasing parameter in terms of the principle squeezing levels and total observed noise magnitude.
\begin{align}
  \Xi &= \frac{\Sigma_-^2}{|\Mq|^2 + |m_b|^2}
\end{align}
The specific formulas \ref{eq:theta_calculation} and \ref{eq:Xi_calculation} follow from the analytic computation of the SVD for 2-by-2 matrices, which generates a specific expression for the singular values, but is too unwieldy to include for the exact angle $\theta(\Omega) \equiv \theta_{\text{D}}$. Instead, an approximation to $\theta_\text{D}$ is given in the limit of small $\Xi$.

\section{Including Phase Noise with Dephasing}
\label{sec:effective_dephasing}
The dephasing parameter $\Xi$ is derived as an intrinsic parameter due only to the optical system; however, it enters the response \cref{eq:metric_S} exactly the same as the non-intrinsic phase noise $\phiRMSsq$ evaluated in \cref{eq:etaSQZ_basic2}. Applying the same expectation operator on \cref{eq:metric_S} as was done for \cref{eq:etaSQZ_basic2}, generates this sequence of squeeze-level parameters $S_{0,1,2}$.
\begin{align}
  S_{0\pm} &= e^{\pm 2r}
             \\
  S_{1\pm} &= (1 - \phiRMSsq) S_{0\pm} + \phiRMSsq  S_{0\mp}
             \\
  S_{2\pm} &= (1 - \Xi)S_{1\pm} + \Xi S_{1\pm}
\end{align}
Which may be expanded and then collected into the effective dephasing $\Xi_{\text{eff}}$.
\begin{align}
  \Xi_{\text{eff}}(\Omega) &= \Xi + \phiRMSsq - 2\Xi\phiRMSsq
\end{align}
This equation maintains the limits that $0 \le \Xi_{\text{eff}} \le 0.5$.

%

\section{Derivations of Passive Transmission Response}
\label{sec:passive_derivations}
The response metrics for passive cavities of \crefrange{eq:theta_passive}{eq:Xi_passive} can certainly be derived using \cref{sec:derivations}, but the passivity constraints provide an alternative derivation. This derivation provides some meaningful insight as it can be done more natively using cavity transfer functions $\tfh(\Omega)$. This work chooses to only represent 2-photon matrices in the quadrature basis of $\hat{q}(\Omega)$ and $\hat{p}(\Omega)$, rather than the sideband basis used for $\hat{a}(\Omega)$ and  $\hat{a}^\dagger(-\Omega)$. One can transform between the two using the $\Tmat{A}$ matrices defined below. For a passive system, $\Tmat{H}$ can be calculated using only $\tfh(\Omega)$, basis-changing \cref{eq:qp_out}, as:
\begin{align}
  \Tmat{H}(\Omega) = \Tmat{A}
  \begin{bmatrix}
    \tfh(+\Omega) & 0 \\
    0 & \conj{\tfh}(-\Omega)
  \end{bmatrix}
        \Tmat{A}^{-1}
\end{align}
\begin{align}
  \Tmat{A} &= \frac{1}{\sqrt{2}}\begin{bmatrix}
    1 & 1\\
    -i & i
  \end{bmatrix}
        &
  \Tmat{A}^{-1} &= \frac{1}{\sqrt{2}}\begin{bmatrix}
    1 & i\\
    1 & -i
  \end{bmatrix}
\end{align}
For a passive system $\Gamma=1$, so \cref{eq:decomposition_relation} simplifies to
\begin{align}
  \eta(\Omega)S(\phi, r) &= \left|\TLO \Tmat{H}\Tmat{R}(\phi)
\Tmat{S}(r)  \right|^2
        \label{eq:var_simple}
\end{align}
When $\tfh(\Omega)$ is reduced by loss, \cref{eq:var_simple} must be extended to include $\Tmat{T}$ terms to couple in un-squeezed vacuum. The passivity condition \cref{eq:passivity_condition} includes every loss source individually accounted, but they can be collected into the complementary loss transfer function $\tfh_{\text{loss}}(\Omega)$.
\begin{align}
  \Hloss &= \Tmat{A}
  \begin{bmatrix}
    \tfh_{\text{loss}}(+\Omega) & 0 \\
    0 & \conj{\tfh}_{\text{loss}}(-\Omega)
  \end{bmatrix}
        \Tmat{A}^{-1}
  \\
  \Tmat{1} &= 
        \Tmat{H}
        \Tmat{H}^\dagger
             +
        \Hloss
        \Hloss^\dagger
\end{align}
The conservation of phase space under the given assumptions
imposes the constraint
\begin{align}
 |\tfh_{\text{loss}}(\pm \Omega)|^2 = 1 - |\tfh(\pm \Omega)|^2
\end{align}
The total noise of \cref{eq:N_no_sqz_derivation} can then be expressed 
\begin{align}
  N &= \TLO\left(  \Tmat{H}\Tmat{R}(\phi)
      \Tmat{S}(r)
      \Tmat{S}^\dagger(r)
        \Tmat{R}(\phi)^{\dagger}\Tmat{H}^{\dagger}
        +
        \Hloss
        \Hloss^\dagger
 \right)
        \TLOa
\end{align}
Together, the efficiency $\eta$ is calculated
\begin{align}
  (1 - \eta) &=
               |
        \TLO
        \Hloss
               |^2
        = 1 - \frac{|\tfh(+\Omega)|^2 + |\tfh(-\Omega)|^2}{2}
\end{align}
Now, for the remaining parameters, some factorizations into magnitude and phase components are needed.
\begin{align}
  \tfh(\pm \Omega)
  &= |\tfh(\pm \Omega)|e^{i \theta_\pm}
  \\
  \Tmat{H}(\Omega)
  &=
    \Tmat{A}
  \begin{bmatrix}
    |\tfh(+\Omega)|e^{i \theta_+} & 0 \\
    0 & |\tfh(-\Omega)|e^{-i \theta_-}
  \end{bmatrix}
    \Tmat{A}^{-1}
  \\
\end{align}
The factorizations then enable an SVD-like decomposition into common and differential magnitudes and phase.
\begin{align}
  C(\Omega) &\equiv \frac{|\tfh(+\Omega)| + |\tfh(-\Omega)|}{2}
      &
\theta_C(\Omega) &\equiv \frac{\theta_+ - \theta_-}{2}
      \\
  D(\Omega) &\equiv \frac{|\tfh(+\Omega)| - |\tfh(-\Omega)|}{2}
           &
\theta_D(\Omega) &\equiv \frac{\theta_+ + \theta_-}{2}
\end{align}
\begin{align}
  \Tmat{H}(\Omega)
           &= 
    \Tmat{A}
  e^{i \theta_C}
  \begin{bmatrix}
    (C + D)e^{i \theta_D} & 0 \\
    0 & (C - D)e^{-i \theta_D}
  \end{bmatrix}
    \Tmat{A}^{-1}
\end{align}
The rotation operator $\Tmat{R}(\phi)$ is a result of phase in the sideband picture, and allows the decomposition to be reduced
\begin{align}
  \Tmat{R}(\phi)
  &=
    \Tmat{A}
    \begin{bmatrix}
      e^{i\phi} & 0\\
      0 & e^{-i\phi}
    \end{bmatrix}
    \Tmat{A}^{-1}
          \label{eq:rotation_identity}
          \\
  \Tmat{H}(\Omega)
           &=
    e^{i \theta_C}\left(
    C\Tmat{R}(\theta_D) + iD\Tmat{R}(\theta_D - \tfrac{\pi}{2})
     \right)
\end{align}

plugging this back into \cref{eq:var_simple} gives
\begin{align}
  N &= \left| \begin{bmatrix}0 \\ 1\end{bmatrix}^\dagger (C\Tmat{1} - D\Tmat{\bbsigma})\Tmat{R}(\theta_D + \phi - \xi) \Tmat{S} \right|^2
        &
  \Tmat{\bbsigma} &= \begin{bmatrix}
    0 & -i \\
    i & 0
    \end{bmatrix}
\end{align}
Using $\xi {-} \sqzang = \theta_D$ for simplicity, this then gives the final phase-quadrature power spectrum of:
\begin{align}
  N(\Omega) &=  C^2 e^{-2r} + D^2 e^{+2r} + (1 - \eta) 
              \\
  &= \eta\left((1 - \Xi) e^{-2r} +  \Xi e^{+2r} \right) + (1 - \eta)
\end{align}
Where the the second line is a result of the following relations:
\begin{align}
\Xi &= D^2
      / \eta \hspace{3em}
  \eta = C^2 + D^2
\end{align}
Relaxing $\xi {-} \sqzang = \theta_D$ can be done to indicate the squeezing angle dependence, but from the above relations, \crefrange{eq:theta_passive}{eq:Xi_passive} follow.

\section{Double Cavity Approximations}
\label{sec:scalar_coupled_cavity}
The transfer function equations \crefrange{eq:u_factors}{eq:double_cavity} are a reduced representation of a double cavity system designed for resonant sideband extraction. Those equations give the reflectivity factorized into roots, zeros and poles, from which analytical expressions can be more easily manipulated. Those roots represent a low order approximation of the response of two cavities, each with differing frequency response. The interaction between the cavities from the common mirror, the arm input mirror, causes a complicated responses that is sensitive to multiple scales of bandwidth, delay time, and resonant enhancement. The reflectivity transfer function of a single transverse mode can be expressed exactly, using:
\begin{align}
  \tfr_\Arm(\Omega)
  &=
    r_\itm  -
    \frac{T_\itm \sqrt{1 - \Lambda_\arm}e^{-i \Omega 2 L_\arm / c + i\psi_\arm}}
    {1 - r_\itm \sqrt{1 - \Lambda_\arm}e^{-i \Omega 2 L_\arm / c + i\psi_\arm}}
    \label{eq:ARM_refl_exact}
    \\
  \tfr_\Src(\Omega)
  &=
    r_\srm  -
    \frac{T_\srm \tfr_\Arm(\Omega)\sqrt{1 - \Lambda_\srm }e^{i \Omega 2 L_\srm  / c + i\psi_\src}}
    {1 - r_\srm \tfr_\Arm(\Omega)\sqrt{1 - \Lambda_\srm }e^{i \Omega 2 L_\srm  / c + i\psi_\src}}
    \label{eq:SRC_refl_exact}
\end{align}
Where $\tfr_\Src$ is the reflectivity of the combined cavity system off of the signal recycling mirror with reflectivity $r_\srm$. $\tfr_\Arm$ is the reflectivity of the arm alone, ignoring the effect of the coupling of the cavities. In LIGO's operating regime, $T_\itm $ is small, to allow a large build up of arm power of the carrier field. $T_\srm $ is large to create a low finesse cavity that only moderately widens the arm bandwidth to be above the frequencies of astrophysical signals. The combination of low and high cavity finesses, as well as the discrepancy in the lengths of the arm and SEC cavities, makes Taylor expansions or Pad\'e approximants of \cref{eq:SRC_refl_exact} nontrivial to construct\cite{Buonanno-PRD03-ScalingLaw}. Furthermore, approximants tend to operate only in a limited parameter regime. To create the approximations used in this work, the following relations are used:
\begin{align}
  T_\itm  &= 1 - \tfr_\itm ^2
  &
  T_\srm  &= 1 - \tfr_\srm ^2
  \\
    \tfr_\itm &= 1 - u_\itm
  &
    \tfr_\srm &= 1 - u_\srm
  \\
  \sqrt{1 - \Lambda_\arm} &\approx 1 - \Lambda_\arm/2
  &
  \sqrt{1 - \Lambda_\src } &\approx 1 - \Lambda_\src /2
\end{align}
The phase shift terms $\Psi_{\{\}}$, which can represent either length detunings or the Gouy phase, are set to 0 for the fundamental mode. The exponential term for the propagation delay is then substituted for a Pad\'e approximant.
\begin{align}
  e^{-i \Omega 2 L / c} &\approx \frac{c - i\Omega L}{c + i\Omega L}
\end{align}
From there, \cref{eq:ARM_refl_exact} is substituted into \cref{eq:SRC_refl_exact} and expanded using computer algebra software. Now, terms are progressively dropped and the transfer function is tested again to the exact one, maintaining the magnitude and phase response as best as possible, even when the loss terms $\Lambda_{\{\}}$ are nonzero. This leads to the following second order rational form.
\begin{align}
  \tfr_\Src(\Omega) &\approx
  \frac{
    a_{2}s^2
    + a_{1}s
    + a_{0}
  }{
    b_{2}s^2
    + b_{1}s
    + b_{0}
  }
        \label{eq:second_order_cavity}
  & s &= i\Omega
  \\
a_{2} &= 2 L_\arm L_\src
         &
a_{0} &= c^2 \left(\textstyle u_\itm  - \frac{\Lambda_\itm u_\srm }{4}\right) 
  \\
b_{2} &= 2 L_\arm L_\src
  &
b_{0} &= c^2 \left(\textstyle u_\itm  + \frac{\Lambda_\itm u_\srm }{4}\right)
         \\
a_{1} &= c L_\arm  \left(\textstyle \frac{\Lambda_\src }{2}(1 - \frac{u_\itm }{2}) - u_\srm \right)\hspace{-5em}
        \label{eq:a1_term}
         \\
b_{1} &= c L_\arm  \left(\textstyle \frac{\Lambda_\src }{2}(1 - u_\srm  - \frac{u_\itm }{4}) + u_\srm \right)\hspace{-5em}
        \label{eq:b1_term}
\end{align}
The rational form is then factored into roots using an approximation of the quadratic formula. Notably, the $a_1$ and $b_1$ terms have different numbers of summed terms, leading to the poles and zeros also having different numbers of terms. By splitting the roots into bandwidth, $\gamma$, and loss, $\lambda$, contributions, the presence of the loss-related terms in the poles and zeros is symmetrized.
\begin{align}
  \gamma_\Arm  - \lambda_\Arm  &\approx -\frac{a_{0}}{a_{1}}
  &
  \gamma_\Arm + \lambda_\Arm &\approx -\frac{b_{0}}{b_{1}}
  \\
  \gamma_\Src - \lambda_\Src &\approx -\frac{a_{1}}{a_{2}} + \frac{a_{0}}{a_{1}}
  &
  \gamma_\Src + \lambda_\Src &\approx -\frac{b_{1}}{b_{2}} + \frac{b_{0}}{b_{1}}
\end{align}
Solving for the individual loss and bandwidth factors for each cavity then leads to \crefrange{eq:u_factors}{eq:delta_factors}, and plugging them back in to \cref{eq:second_order_cavity} leads to \cref{eq:double_cavity}.

\section{Multiple Transverse Mode Interferometer Model}
\label{sec:matrix_coupled_cavity}
The effort of this paper is primarily to produce simplified models the squeezing response in light of transverse mismatch and radiation pressure. To validate those models, it is useful to compare against a more complete, though opaque, model that includes the exact cavity response with radiation pressure, detuning, losses, and transverse modal mismatch. With the exception of transverse mismatch, such a model is established and widely used for noise modeling of for LIGO-like interferometers\cite{Kimble-PRD01-ConversionConventional, Buonanno-PRD01-QuantumNoise}. This model is succinctly derived here in a manner that allows transverse mismatch to be incorporated. For simplicity, and to provide a more direct comparison, this is done for a double cavity representing a perfectly symmetric interferometer. Future work should simulate interferometers with arm imbalances to compare against this exact case, but that is beyond the scope chosen here.

To incorporate all of the listed elements in an exact model, a product space is necessary to maintain the two-photon response of each optical element across multiple inter-coupled transverse modes. Here, only two such transverse modes are used, the fundamental Gaussian mode and a single higher order mode such as the LG1 for beam mismatch, or the HG01 for a misalignment. The interaction between the modes conserves the phase space and does not leak into yet higher modes. Similarly to \cref{eq:U_def}, this is not a perfectly physical choice, but a convenient one that is valid for small mixing between the modes. The squeezing, rotation, and mode mixing matrices in this product space are defined below in terms of their two-photon definitions. The rotation matrix takes on two parameters, one common rotation $\phi$, representing a phase shift of both modes, and one $\psi$ for the rotation solely of the higher order mode (HOM).
\begin{align}
\Dmat{S}(r) &\equiv
\begin{bmatrix} 
\Tmat{S}(r) & \Tmat{0} \\
\Tmat{0} & \Tmat{1}
\end{bmatrix}
&
\Dmat{R}(\phi, \psi) &\equiv
\begin{bmatrix} 
\Tmat{R}(\phi) & \Tmat{0} \\
\Tmat{0} & \Tmat{R}(\psi)\Tmat{R}(\phi) 
\end{bmatrix}
\end{align}
The mismatch loss coupling matrix $\Dmat{U}$ maintains the same parameters as before \cref{eq:U_def}. The HOM phase shift term must be converted into a quadrature rotation, and the common phase $\theta$ is omitted.
\begin{align}
\Dmat{U}(\MM, \psi) &\equiv
  \begin{bmatrix}
    \sqrt{1 - \Upsilon}\Tmat{1} & -\sqrt{\Upsilon}\Tmat{R}(\psi)
    \\
    \sqrt{\Upsilon}\Tmat{R}(-\psi) & \sqrt{1 - \Upsilon}\Tmat{1}
  \end{bmatrix}
\end{align}
Additionally, basis vectors in this space are defined to simplify the expression of single-element matrices as well as create projections for observables.
\begin{align}
    \begin{bmatrix}
    \Dvec{e}_{q0} &
    \Dvec{e}_{p0} &
    \Dvec{e}_{q1} &
    \Dvec{e}_{p1}
  \end{bmatrix}
  &\equiv 
  \Dmat{1}
\end{align}
As in the scalar transfer function case of \cref{sec:scalar_coupled_cavity}, the reflectivity of the interferometer double cavity system from the signal-recycling mirror is needed. To make analogous equations to \cref{eq:ARM_refl_exact} and \cref{eq:SRC_refl_exact}, most of the same scalar factors are needed, but now in the product space.
\begin{align}
  \Dmat{r}_\itm &= \sqrt{1 - T_\itm}\cdot\Dmat{1}
                  &
  \Dmat{t}_\itm &= \sqrt{T_\itm}\cdot\Dmat{1}
  \\
  \Dmat{r}_\etm &= \sqrt{1 - T_\etm}\cdot\Dmat{1}
                  &
  \Dmat{t}_\etm &= \sqrt{T_\etm}\cdot\Dmat{1}
  \\
  \Dmat{r}_\srm &= \sqrt{1 - T_\srm}\cdot\Dmat{1}
                  &
  \Dmat{t}_\srm &= \sqrt{T_\srm}\cdot\Dmat{1}
  \\
  \Dmat{\eta}_\arm &= \sqrt{1 - \Lambda_\arm}\cdot\Dmat{1}
                  &
  \Dmat{\eta}_\src &= \sqrt{1 - \Lambda_\src}\cdot\Dmat{1}
  \\
  \Dmat{\delay}_\arm &= e^{i\Omega L_\arm / c}\Dmat{R}(k\Delta L_\arm, \psi_\arm)
  &
  \Dmat{\delay}_\src &= e^{i\Omega L_\src / c }\Dmat{R}(k\Delta L_\src, \psi_\src)
\end{align}
The transmission delay matrices $\Dmat{L}_\arm$ and $\Dmat{L}_\src$ use the identity \cref{eq:rotation_identity} in this larger space. For them, the HOM picks up the single pass Gouy phase of the arm $\psi_\arm$ and of the SEC $\psi_\src$. The $\Delta L_\arm$ and  $\Delta L_\arm$ are microscopic detuning lengths for each cavity.

The final new component of this double cavity matrix model is the radiation pressure. This is added ad-hoc as a modification to the reflectivity of each of the arm cavity mirrors. It couples the amplitude and phase quadratures only in the fundamental mode, as that is the mode that the large carrier power $P_\Arm$ resonates in each arm. The modified reflectivities are $\Dmat{\rho}_\itm$ and  $\Dmat{\rho}_\etm$ for the input and end mirror respectively.
\begin{align}
  \Dmat{\rho}_\itm &= \Dmat{r}_\itm  + 8k\cdot \chi(\Omega)\cdot\frac{P_\Arm}{c}\cdot\Dvec{e}_{p0}\Dvec{e}_{q0}^{\dagger}
  \\
  \Dmat{\rho}_\etm &= \Dmat{r}_\etm + 8k\cdot \chi(\Omega)\cdot\frac{P_\Arm}{c}\cdot\Dvec{e}_{p0}\Dvec{e}_{q0}^{\dagger}
\end{align}
The mode mismatch of this model can be added not only between the external elements of the squeezing, interferometer and readout as shown in \cref{fig:SQZ_mm_chain}, but also within the interferometer. The input and output mismatch matrices of \cref{sec:modeling_TMM} are given by $\Dmat{U}_\subI$ and $\Dmat{U}_\subO$, while the mismatch between the signal recycling cavity and the arm cavity is given by $\Dmat{U}_\Arm$. These matrices act as basis transformations into and out of the respective component basis, where the mismatch loss of the fundamental is given by an $\MM_{\{\}}$ parameter, and there is a (generally unknown) mismatch phasing $\psi_{\{\}}$.
\begin{align}
\Dmat{U}_\Arm &= \Dmat{U}(\Upsilon_A, \psi_A)
                       &
\Dmat{U}_\subI &= \Dmat{U}(\MMI, \psi_\subI)
                       &
\Dmat{U}_\subO &= \Dmat{U}(\MMO, \psi_\subO)
\end{align}
The following \crefrange{eq:product_arm_prop}{eq:product_src_refl} are the extensions of \cref{eq:ARM_refl_exact} and \cref{eq:SRC_refl_exact} into the product space. It is solved using noncommutative Gaussian elimination first on the arm, then on the signal recycling cavity. $\Dmat{F}_\Arm$ is the round-trip closed-loop propagator from the end-mirror back to itself via the input mirror. As such, it enters at a specific point -- immediately after the end mirror reflectivity $\Dmat{\rho}_\etm$ -- in the round-trip propagation sequence of the arm cavity reflectivity $\Dmat{r}_\Arm$. Given the placements of the $\Dmat{U}_\Arm$ factors, the arm cavity reflectivity is in the modal basis of the signal recycling cavity.
\begin{align}
  \Dmat{F}_\Arm &=
    \left(
    \Dmat{1} -
    \Dmat{\eta}_\arm 
    \Dmat{\rho}_\etm
    \Dmat{\delay}_\arm
    \Dmat{\rho}_\itm
    \Dmat{\delay}_\arm
    \right)^{-1}
                  \label{eq:product_arm_prop}
  \\
  \Dmat{r}_\Arm
  &=
    \Dmat{U}_A^{-1}
    \left( 
    \Dmat{r}_\itm - 
    \Dmat{t}_\itm
    \Dmat{\delay}_\arm
    \Dmat{F}_\Arm
    \Dmat{\eta}_\arm 
    \Dmat{\rho}_\etm
    \Dmat{\delay}_\arm
    \Dmat{t}_\itm
    \right)
    \Dmat{U}_A
    \label{eq:product_arm_refl}
    \\
  \Dmat{F}_\Src &= 
    \left( 
    \Dmat{1} -
    \Dmat{\eta}_\src
    \Dmat{r}_\Arm
    \Dmat{\delay}_\src
    \Dmat{r}_\srm
    \Dmat{\delay}_\src
    \right)^{-1}
    \label{eq:product_src_prop}
  \\
  \Dmat{r}_\Src
  &=
    \Dmat{U}_\subI^{-1}
    \left(
    \Dmat{r}_\srm - 
    \Dmat{t}_\srm
    \Dmat{\delay}_\src
    \Dmat{F}_\Src
    \Dmat{\eta}_\src
    \Dmat{r}_\Arm
    \Dmat{\delay}_\src
    \Dmat{t}_\srm
    \right)
    \Dmat{U}_\subI
    \label{eq:product_src_refl}
\end{align}
The propagator and reflectivity of the signal recycling cavity are constructed similarly to the arm and, like \cref{eq:SRC_refl_exact}, use the arm cavity reflectivity instead of the arm input mirror reflectivity. This follows from the particular ordering chosen during Gaussian elimination. The placements of $\Dmat{U}_\subI$ indicate that $\Dmat{r}_\Src$ is in the basis of the squeezer input beam.

Along with the reflectivity, all of the $\Dmat{H}$ propagation and $\{\Dmat{T}\}$ loss matrices of \cref{sec:derivations} must be constructed. For brevity, the broadband input and output losses from $\Lambda_\subI$ and $\Lambda_\subO$ are not included, but are simple to incorporate. Instead, the internal interferometer losses from the arm and signal recycling cavities are calculated using the transmission matrices
\begin{align}
  \Dmat{T}_\text{R,\Src}
  &= 
  \Dmat{t}_\Src\sqrt{\Lambda_\src}
  &
  \Dmat{t}_\Src
  &=
    \Dmat{U}_\subI^{-1}
    \Dmat{t}_\srm
    \Dmat{\delay}_\src
    \Dmat{F}_\Src
    \Dmat{\eta}_\src
  \\
  \Dmat{T}_\text{R,\Arm}
  &= 
    \Dmat{t}_\Arm\sqrt{T_\etm + \Lambda_\arm}
  &
    \Dmat{t}_\Arm
  &=
    \Dmat{t}_\Src
    \Dmat{U}_A^{-1}
    \Dmat{t}_\itm
    \Dmat{\delay}_\arm
    \Dmat{F}_\Arm
    \Dmat{\eta}_\arm 
\end{align}
The output mode cleaner is applied to the reflectivity to create the total propagation of squeezing $\Dmat{H}$. Because of the output mode cleaner, the homodyne readout projects solely in the fundamental mode, while the LO angle $\xi \approx 0$ has some freedom due to contrast defect in the interferometer. Together, these generate the effective observation vector $\vec{m}$, which retains only 2 elements and is directly applicable to the formulas of \cref{sec:derivations}.
\begin{align}
  \Dmat{H} &= \Dmat{U}_\subO \Dmat{r}_\Src
  &
  \Mq &= \DLO\Dmat{H}\Dvec{e}_{q0}
  \\
    \DLO &= \Dvec{e}_{p0}^\dagger\Dmat{R}(\zeta, 0)
        &
  \Mp &= \DLO\Dmat{H}\Dvec{e}_{p0}
\end{align}
While $\vec{m}$ still has only 2 elements, the vectors in the norms for the total noise gain are still in the product space, here dimension 4. This sums the higher order mode terms as contributors to loss, decreasing the efficiency in the computation $\eta = |\vec{m}|^2/\Gamma$.
\begin{align}
  \Gamma(\Omega) &= \left|\DLO\Dmat{H}\right|^2
                   + \left|\DLO\Dmat{T}_\text{R,\arm}\right|^2
                   + \left|\DLO\Dmat{T}_\text{R,\src}\right|^2
\end{align}
From the previous expressions, the noise $N(\Omega)$ can be entirely calculated, so squeezing can be examined. With all the machinery, it is also useful to determine the optical gain and signal sensitivity. Below, $\Dvec{S}$ is the signal field generated by displacement modulations $x(\Omega)$. Displacements create phase modulations in the fundamental mode at the end mirror. The factor of $1/\sqrt{2}$ is from the presence of the beamsplitter. The field magnitude in the transverse mode and quadrature observed by the homodyne is given by $S(\Omega)$.
\begin{align}
  S(\Omega) &= \DLO\Dvec{S},
              &
  \Dvec{S} &= 
   \frac{1}{\sqrt{2}}\Dmat{U}_\subO \Dmat{t}_\Arm
  \Dvec{e}_{p0}
             \cdot
  2k\sqrt{\frac{P_\Arm}{\hbar \omega}}
  x(\Omega)
\end{align}
The signal sensitivity can then be used to define the optical gain and sensitivity in terms of spectral density as per \cref{sec:ideal_IFO_metrics}, \cref{eq:optical_calibration_G}. Since $s$ and $\Dvec{s}$ are in units of quanta/second, rather than unitful power, the factor of $1/2$ in $G$ represents the half quanta of vacuum noise.
\begin{align}
  g(\Omega) &= \DLO \Dmat{U}_\subO \Dmat{t}_\Arm \Dvec{e}_{p0},
  &
    G(\Omega) &= \frac{1}{2} \left(L_\Arm \frac{\text{d} S(\Omega)}{\text{d} x(\Omega)} \right)^{-2}
\end{align}

\section{Radiation Pressure Region}
\label{sec:radiation_pressure_calc}

The previous appendix \ref{sec:matrix_coupled_cavity} derived the interferometer squeezing and signal response in full generality. Due to its generality, the full expressions obscure the physics of the ideal and nearly ideal cases. This appendix specifically investigates the interaction of external mode mismatch on the QRPN, where the interferometer itself is lossless and perfectly on resonance. The coherent effects of modal mismatch and the coherent effects in QRPN could provide an alternative explanation of the variation of $\Gamma$ required to explain the loss measurement in the LLO data.

This section concludes that is not the case, and that the LO angle $\xi \ne 0$ variation is a more valid explanation. This derivation also indicates some limitations in using squeezing and QRPN as a diagnostic of the arm power, as the observed $\Gamma(\Omega)$ and $\theta(\Omega)$ do have some dependence on $\MMI$ and $\MMO$, and their dependence mimics lower power in measurements.

To model the ideal interferometer with mismatch, the matrices of \cref{sec:ideal_IFO_example} simply need to be extended into the product space of \cref{sec:matrix_coupled_cavity} to incorporate additional transverse modes. The reflectivity matrix $\Dmat{R}_\Src$ is naturally in a lower diagonal form when the interferometer is on-resonance and has no mismatch, as all of the matrices entering in \crefrange{eq:product_arm_prop}{eq:product_src_refl} are either diagonal or lower triangular, so the matrix inverses simplify greatly, as each diagonal in the inverse becomes a formula like \cref{eq:SRC_refl_exact}. In the inverses, the off diagonal term for the radiation pressure interaction picks up the optical gain to become \cref{eq:optical_gain_g}. Together, the ideal interferometer reflectivity for the squeezing field becomes:
\begin{align}
  \Dmat{r}_\Src(\Omega)
  &= 
    \begin{bmatrix}
      \tfr_2(\Omega) & 0 & 0 & 0
      \\
      \K(\Omega) & \tfr_2(\Omega) & 0 & 0
      \\
      0 & 0 & 1 & 0
      \\
      0 & 0 & 0 & 1
    \end{bmatrix}
\end{align}
Mode mismatch is then added similarly to \cref{sec:modeling_TMM}, except using this product space representation.
\begin{align}
  \Dmat{H}_\subR(\Omega)
  &= 
    \Dmat{U}(\MMO, \psiO)
    \Dmat{G}
    \Dmat{U}^\dagger(\MMI, \psiI)
    \Dmat{r}_\Src
    \Dmat{U}(\MMI, \psiI)
\end{align}

The resulting expressions are complicated, as all the interactions are coherent. Here, row of $\Dmat{H}_\subR$ corresponding to the phase quadrature of the fundamental mode is presented, as that is the only relevant output to calculate squeezing in the ideal case, using phase quadrature readout $\TLO = \Tvec{e}_{p0}^{\dagger}$.
\begin{align}
  \Dvec{e}_{p0}^\dagger\Dmat{H}_{\text{R}}(\Omega)\Dvec{e}_{q0}
    &= \K\sqrt{\cMMO}\left(\cMMI  + c\sqrt{\MMI \MMO}\cos(\psi) \right)
      \nonumber\\
     &= \K\sqrt{\cMMO}\left(1 - \frac{\MMR}{4}\right)
  \\
  \Dvec{e}_{p0}^\dagger\Dmat{H}_{\text{R}}(\Omega)\Dvec{e}_{p0}
  &= \sqrt{\cMMO}
  \\
  \Dvec{e}_{p0}^\dagger\Dmat{H}_{\text{R}}(\Omega)\Dvec{e}_{q1}
    &= -\K\sqrt{\MMI}\sqrt{\cMMI}\sqrt{\cMMO}
      \nonumber\\&\hspace{2em}
  - \K \MMI\sqrt{\MMO}\cos(\psi) - \sqrt{\MMO}\sin(\psi)
  \\
  \Dvec{e}_{p0}^\dagger\Dmat{H}_{\text{R}}(\Omega)\Dvec{e}_{p1}
  &= -\sqrt{\MMO} \cos(\psi)
\end{align}
These 4 terms can be separated into the purely two-photon matrix form of \cref{sec:derivations}, appearing as:
\begin{align}
    \begin{bmatrix}
      \Tmat{H}_\subR & \Tmat{\Lambda}_\subR
      \\
      \cdot & \cdot
    \end{bmatrix}
  &\equiv
  \Dmat{H}_\subR(\Omega)
  & 
\end{align}
These can then be used to compute the squeezing response metrics:
\begin{align}
  \eta(\Omega) \Gamma(\Omega)
  &=
  (\cMMO)\left(1 + \K^2\left(1 - \frac{\MMR}{2}\right)\right)
  \\
  \Gamma(\Omega)
  &= \left| \Dvec{e}_{p0}^\dagger\Dmat{H}_{\text{R}}(\Omega) \right|^2 = 1 + \K^2(\cMMF)
  \\
  \eta_\subR(\Omega)
  &
    =  (\cMMO)\frac{1 + \K^2\left(1 - \MMR/2\right)}{1 + \K^2(\cMMF)}
  \\
  \theta_\subR(\Omega) &= \arctan\left(|\K|\left(1 - \frac{\MMR}{4}\right)\right)
\end{align}
Notably, the transverse mismatch losses entail certain adjustments needed to the noise gain $\Gamma(\Omega)$ and observation angle $\theta(\Omega)$. The $\K$ term in $\Gamma$ is diminished by the readout losses in exactly the same manner that the signal experiences. This corresponds to how the quantum noise in amplitude causes force and displacement that mimics signals, so the quantum noise gain must be enhanced or reduced to the same degree.

$\theta(\Omega)$ behaves differently, as it reacts to both interactions which coherent modify it at once. The influence of mismatch losses bias its estimate of $\K$, and thus the arm power, downwards. This happens because the mode mismatch moves some amount of squeezing out of the interferometer mode, preventing it from experiencing $\K$, then moves it back. This effectively causes $\K$ to appear reduced.

The efficiency $\eta_\subR$ is also affected by the mismatch loss and has a slightly different adjustment of its $\K$ terms between the numerator and denominator. This causes the orange model curves in the middle plots of loss in \cref{fig:data_Q} to wiggle downwards at low frequencies. This wiggle doesn't have particular physical significance, and can be interpreted as evidence that $\eta \Gamma$ is a more fundamentally useful metric than $\eta$ or $\Gamma$ alone. Notably,  $\eta \Gamma$ has the same dependence on mismatch from $\MMR$, per factor of $\K$, as $\theta$.

\end{document}